\documentclass{emulateapj}

\usepackage{graphicx,amsmath}
\newcommand{\edII}[1]{}

\begin{document}

\title{Gas emission from debris disks around A and F stars}

\author{Kyryl Zagorovsky}
\affil{IBBME, University of Toronto, Toronto, ON M5S\,3G9, Canada}
\email{kyrylz@gmail.com}

\author{Alexis Brandeker}
\affil{Department of Astronomy, Stockholm University, SE-106\,91 Stockholm, Sweden}
\email{alexis@astro.su.se}

\author{Yanqin Wu}
\affil{Department of Astronomy and Astrophysics, University of Toronto, Toronto, ON M5S\,3H4, Canada}
\email{wu@astro.utoronto.ca}

\begin{abstract}

Gas has been detected in a number of debris disk systems. This gas may
have arisen from grain sublimation or grain photodesorption. It
interacts with the surrounding dust grains through a number of charge
and heat exchanges.  Studying the chemical composition and physical
state of this gas can therefore reveal much about the dust component
in these debris disks.  We have produced a new code,
\textsc{ontario}, to address gas emission from dusty gas-poor 
disks around A--F stars.  This code computes the gas ionization and
thermal balance self-consistently, with particular care taken of
heating/cooling mechanisms. Line emission spectra are then produced
for each species (up to zinc) by statistical equilibrium calculations
of the atomic/ionic energy levels.
For parameters that resemble the observed $\beta$~Pictoris gas disk, we
find that the gas is primarily heated by photoelectric emission from
dust grains, and primarily cooled through the \ion{C}{2} 157.7\,$\mu$m
line emission. The gas can be heated to a temperature that is warmer
than that of the dust and may in some cases reach temperature for
thermal escape.  The dominant cooling line,
\ion{C}{2} 157.7\,$\mu$m, should be detectable by
\textit{Herschel} \edII{in these disks}, while the \ion{O}{1} 63.2\,$\mu$m line will
be too faint. We also study the dependence of the cooling line fluxes on a
variety of disk parameters, in light of the much improved
sensitivity to thermal line emission in the mid/far infrared and at 
sub-millimeter wavelengths provided by, in particular, 
\textit{Herschel}, \textit{SOFIA}, and \textit{ALMA}. These new 
instruments will yield much new information about dusty debris disks.

\end{abstract}



\section{Introduction}

During the T~Tauri phase, models have shown that giant planet
formation rely heavily on the timescale and manner in which gas disks
disperse \citep[see, e.g.][]{idalin}. We now know that some of this
gas is incorporated into gaseous giant planets, while some is lost to
the stars, and that most of the gas disappears on timescales of a
few Myr \citep[e.g.][]{zuc95,hai01,jay06,pas06}. 
However, we still do not know whether most of the gas disks around
T~Tauri stars is photoevaporated by the star
\citep[see, e.g.][]{gor08}, blown away by wind \citep{lovelace}, or accreted onto
the star due to viscous diffusion \citep{lyndenbell}. Which mechanism
dominates will determine how long gas resides in different parts of
the disk. Observing protoplanetary disks near their final stages may
yield clue to this puzzle.

Debris disks are circumstellar disks that show emission from
dust. These disks are detectable for up to a few Gyr
\citep{hol98,spa01,rie05}.
Both the level of emission and the fraction of stars with a detectable
excess decay with time and can be interpreted, at least for the case
of A stars, as steady state collisional processing of planetesimals
formed during the proto-planetary disk phase
\citep{dom03,wya07}. There is some evidence 
that gas may persist in debris disks for an extended period of time,
at least in some systems. For example, atomic gas in Keplerian
rotation has been detected around the $\sim 12$\,Myr old main-sequence star 
$\beta$\,Pic \citep{hob85,bra04}. Gas has also been
found around post-T~Tauri stars \citep[e.g.][]{red07b}, and
even stars as old as 200\,Myr \citep[e.g.\ $\sigma$\,Her;][]{che03}. 
However, it is not clear whether this
gas is primordial (left-over proto-planetary material) or secondary,
generated by, e.g., sublimation of planetesimals
\citep{beu07}, vaporization of colliding planetesimals
\citep{cze07}, and/or photo-desorption of dust
\citep{che07}. If gas is indeed 
secondary \citep*[as dynamical arguments suggest, see][]{fer06}, it
presents an exciting new tool to study the compositions of solid
bodies in extra-solar systems, unaccessible otherwise. In addition,
even a small quantity of gas may also significantly affect the
dynamics of the dust and cause grain concentrations, mimicking the
bright rings seen in young debris disks like HR\,4796A
\citep{kla05,bes07}. Furthermore, gas present during late stages of 
planet formation could damp eccentricities of the planetesimals enough 
to reduce their relative speed, thereby aiding the build up of rocky planets.

Unfortunately, the information currently available on the manner in
which gas disperses in circumstellar disks is scarce, owing to the
great difficulty in detecting gas in these environments.  Accretion
traces only the sub-AU region of the disk, leaving the outer gas disk
largely unconstrained.  Molecular emission from CO in young
circumstellar disks has been detected in the near-infrared (NIR) and
in the mm \citep[e.g.][]{naj03}. The NIR emission is limited to hot CO
($\sim 1\,000$\,K), while the detection of mm emission from cold CO is
hampered by limited sensitivity due to beam dilution
\citep[e.g.][]{pas06}.  Direct detection of H$_2$ in the
ultra violet (UV) and NIR is also made possible by either very hot or fluorescent gas
\citep[pumped by Ly\,$\alpha$, ][]{her06}. The rotational lines H$_2$ S(0) and S(1)
emitted from colder gas ($\lesssim 100$\,K) are unfortunately weak and
difficult to observe: a Spitzer IRS survey of young stars that are
expected to have gas disks yielded only upper limits on the gas disk
mass \citep[upper limits $\gtrsim100\,M_{\oplus}$,][]{che06}.  An
alternative is to observe stellar light scattered resonantly by the
gas, but this requires a spatially resolved disk to be detectable. To
date, only the disk around $\beta$\,Pic has been detected in such a
way \citep{olo01,bra04}. 
Should the disk be oriented edge-on, the circumstellar gas can also be
observed in absorption \citep[e.g.,][]{red07a,rob08}.

Here, we study infrared atomic cooling lines as promising agents for
characterizing gas in tenuous disks, including late stage
protoplanetary disks and debris disks. Three of the most abundant
species, oxygen, carbon, and silicon, all have ground-state fine
transitions at infrared wavelengths that act as effective coolants.
Oxygen, in the low-density environment of these disks, is likely to be
in the atomic form: the molecular repositories for oxygen
\citep[H$_2$O, OH, and CO; e.g.][]{kam00} are dissociated easily in
the circumstellar environment, while the ionization potential for
oxygen is sufficiently high for it to remain neutral.  Moreover,
neutral oxygen does not have strong resonant transitions in the
optical and UV, and is thus not easily removed by radiation pressure,
unlike some other species \citep{fer06}. Of all thermally excited
cooling lines, the \ion{O}{1} 63\,$\mu$m line is expected to be the
brightest for a large range of disk masses in late-type stars
\citep[G--K;][]{gor04}. However, in optically thin disks around
early-type stars (F or earlier), carbon is expected to be
significantly ionized \citep{fer06}, boosting the \ion{C}{2}
158\,$\mu$m cooling line. Finally, if cooling by the \ion{O}{1} and
\ion{C}{2} lines saturates (i.e.\ the line fluxes no longer rise
  with temperature due to the saturated occupation of the excited
  level, see eq.~[\ref{eq:ninx}]), \ion{Si}{2}-cooling may become
important.

The \ion{O}{1} 63\,$\mu$m and \ion{C}{2} 158\,$\mu$m lines both lie
within the spectral range of the far-infrared to sub-millimeter 
\textit{Herschel} space telescope, scheduled to start science operations 
in early 2010. The telescope has a passively cooled 3.5\,m mirror and three
instruments PACS, SPIRE, and HIFI that together cover the 55--672\,$\mu$m
with both imaging and spectroscopy capabilities. As such, Herschel is well 
positioned to investigate the tenuous circumstellar disks we are 
discussing here and partly motivated this study.

Predicting the expected luminosity of gas and interpreting potential
detections require detailed modeling of the physical circumstances,
such as the radiation field from the star, the spatial distribution
and content of the gas and dust, etc.  Debris disks are known to be
dusty with little gas, meaning that the set of assumptions and
approximations that can be used are different from other environments,
such as the interstellar medium where well-tested codes exist
\citep[e.g.\ \textsc{cloudy},][]{fer05}. Debris disks are generally
optically thin, which simplifies the radiative transfer and enables
modeling of more general three-dimensional distributions with less
computational resources. On the other hand, a more detailed treatment
is required for the thermal balance, as fluorescence is generally
important due to the strong radiation field from the star in
combination with low gas densities. This implies computing the
statistical equilibrium (SE) for a large number of energy levels in
cooling species.

We have developed a numerical model \textsc{ontario} (``optically thin thermal and ionization 
equilibrium code'') to
investigate the importance of \ion{O}{1} 63\,$\mu$m and \ion{C}{2} 158\,$\mu$m
fine-structure cooling lines as tracers for gas in dusty but optically
thin disks around early-type (A--F) stars. This is intended to complement
similar analysis made for later-type stars (G--K) by \citet{gor04}. The
code takes stellar flux, gas and dust profiles as input parameters and
performs self-consistent computation of ionization and thermal
balance (\S\,\ref{sec:thermal}) in the disk. Once elemental
ionizations and gas temperature have been determined, a full SE 
computation is performed for a number of atomic
species to identify the dominant emission and absorption lines.

To test the code, we use the well-studied 
$\beta$~Pictoris disk as a benchmark case in
\S\,\ref{sec:BPic}, and compare the 
numerical model with observed gas disk properties.
In \S\,\ref{sec:Fiducial}, we keep stellar parameters at
their $\beta$\,Pic values but replace the gas and dust distributions with
more simplified Gaussian profiles, and take this configuration as the
fiducial debris disk. Fiducial case temperature and ionization
profiles are analyzed in \S\,\ref{sec:Fid_Temp} and
\S\,\ref{sec:Fid_Ion}. We also compute integrated luminosities for the
main fine-structure cooling lines (\S\,\ref{sec:Fid_Cool}) and identify
the dominant emission and absorption lines
(\S\,\ref{sec:Fid_Spectr}). In \S\,\ref{sec:Discussion} we investigate
the dependence of emitted line luminosities on the model parameters, in
particular how they scale with disk gas, geometry, mass and composition,
and the spectral type of the central star. Conclusions are drawn in 
\S\,\ref{sec:Conclusion}.

Throughout the paper, we adopt the cgs units except where noted.

\section{Thermal processes modeled in \textsc{ontario}}
\label{sec:thermal}

The \textsc{ontario} model was designed to simulate gas in dusty disks around
early-type stars. The code self-consistently computes the thermal and 
ionization states of the gas in this environment.  All
gas is assumed to be present in atomic or ionized form: the hard radiation
fields surrounding these early-type stars rapidly photo-dissociate molecules
into individual atoms or ions \citep{kam03,jon06}; thus no chemistry computation 
is included. Ionizations for atomic elements from hydrogen to zinc are
considered, up to the second ionization state. The model assumes both 
dust and gas to be optically thin, so no complex radiative transfer 
calculations are performed. This makes the code effectively zero-dimensional
as each bin is treated independently of all others, greatly increasing 
computational performance. Details of the code, including how ionization
and statistical equilibrium is treated, is presented in the Appendix. Here, we
discuss the main processes that enter the thermal balance.

\defcitealias{kam01}{KvZ01} 
\defcitealias{bes07}{BW07}
A number of heating and cooling mechanisms that determine the gas
temperature in debris disks have been investigated by
\citet[ hereafter \citetalias{kam01}]{kam01} and
\citet[ hereafter \citetalias{bes07}]{bes07}.
 Observed upper limits on the gas content of debris disks
\citep{thi01,che03,rob05} and theoretical inferences
\citep{fer06} suggest that gas in debris disk is 
non-primordial and is depleted in hydrogen. Furthermore, even when
assuming solar abundance of hydrogen, it was shown by
\citetalias{bes07} that heating arising from H$_2$ collisional
de-excitation, photodissociation and formation on dust, as well as
cooling from H$_2$ ro-vibrational lines, are negligible when compared
to the photoelectric heating by grains. For these reasons, we do not
include H$_2$-dependent heating and cooling processes. We do include,
however, the atomic hydrogen contribution to cooling via Ly\,$\alpha$
emission at very high temperatures ($T > 5\,000$\,K).
Also excluded from computation are coolings by vibrational/rotational
transitions of molecular CO and CH. These molecular species
may arise from evaporating cometary bodies. However, they are
  believed to be quickly photo-dissociated around the early-type stars
  that are the target of this study \citep{kam03}. Indeed,
\textit{HST} UV observations of $\beta$\,Pic failed to detect
significant amounts of CO in the system \citep{rob00}.

The gas temperature is solved at every spatial grid assuming that heating
and cooling processes balance each other locally.

\subsection{Heating}
\label{sec:heating}
Relevant heating processes include: photoelectric heating by dust
(PE), photoionization of gas by the stellar light (PI), and gas-grain
collisions (GG). The last one becomes a cooling mechanism for the gas
when $T_{\mathrm{gas}} > T_{\mathrm{dust}}$. For $\beta$\,Pic-like disks, 
\citetalias{bes07} have shown that photoelectric (PE) heating is by far 
the dominant heating mechanism.
\paragraph{Photoelectric heating.}
PE is caused by energetic stellar photons striking dust particles, ejecting 
electrons from the grains. Released electrons, in turn, contribute their
kinetic energy to the gas. The PE heating rate per unit volume is given by
\citet{wei01}\footnote{The
\citet{wei01} formulation is a refinement over that in \citet{dra78}. 
We adopt it here. } as
\begin{eqnarray*}
  \Gamma_{\mathrm{PE}} = \int_{s_{\mathrm{min}}}^{s_{\mathrm{max}}} ds
  \frac{dn_{\mathrm{dust}}}{ds} \pi s^2 \int^{\nu_{\mathrm{max}}}_{(e\phi+W)/h}
  Q_{\mathrm{abs}}Y(h\nu) \frac{F_{\nu}}{h\nu}d\nu
\end{eqnarray*}
\begin{equation}
  \times \left[\int^{h\nu-W-e\phi}_{0}Ef(E,h\nu)dE\right].
  \label{eqn:Gampe}
\end{equation}
Here, $\int_{s_{\mathrm{min}}}^{s_{\mathrm{max}}} ds \frac{dn_{\mathrm{dust}}}{ds} \pi s^2$ 
represents the total dust area per unit
volume, coming from grains of radii between $s_\mathrm{min}$ and
  $s_\mathrm{max}$, that stellar photons can intercept.  $\phi$ is
the charging potential of the grain, $W$ is its work function,
$Q_{\mathrm{abs}}$ is the overall absorption coefficient, $Y(h\nu)$ is
the photoelectric yield, $F_{\nu}$ is the stellar flux at a given
frequency, $E$ is energy of the ejected electrons, and $f(E,h\nu)$
describes their energy distribution.  Following \citet{wei01}, we take
$Q_{\mathrm{abs}} = 1$.  Throughout this paper we assume a
carbonaceous composition for the dust grains, giving a work function
of $W = 4.4$\,eV. Also,
\begin{eqnarray}
Y(h\nu) & = & {\frac{E_{\mathrm{high}}^2 (E_{\mathrm{high}} - 3 E_{\mathrm{low}})}{
(E_{\mathrm{high}} - E_{\mathrm{low}})^3}} \times y_0, \nonumber \\
f(E,h\nu) & = & {\frac{6(E-E_{\mathrm{low}}) (E_{\mathrm{high}} - E)}{E_{\mathrm{high}}^2
(E_{\mathrm{high}} - 3 E_{\mathrm{low}})}},
\end{eqnarray}
where $E_{\mathrm{high}} = h\nu - W - e\phi$, $E_{\mathrm{low}} = -
e\phi$ and the dimensionless factor $y_0$ is the highly uncertain
photoionization yield. Following \citet{wei01}, we take
\begin{eqnarray}
y_0 & = & {\frac{9\times 10^{-3} (h\nu - W)^5}{W^5 + 3.7\times 10^{-2} 
(h\nu - W)^5}} \hskip0.6in {\mathrm{carbonaceous}}, \nonumber \\
    & = &  \frac{0.5 (h\nu - W)}{W + 5 (h\nu-W)} \hskip1.2in{\mathrm{silicate}}.
\end{eqnarray}
Each photoelectric electron that makes it out of the grain carries
with it an average energy of order $1$\,eV. 

The charging potential $\phi$ critically determines the heating
capability of stellar photons -- only stellar photons with energies
above a threshold ($W + e\phi$) yield PE heating.  The potential is
independent of grain size and is obtained by equating the
photoelectric charging current per unit area on the dust grain,
\begin{equation}
  J_{\mathrm{PE}} =  \frac{e}{4}
\int^{\nu_{\mathrm{max}}}_{(e\phi+W)/h}
  Q_{\mathrm{abs}}Y(h\nu) \frac{F_{\nu}}{h\nu}d\nu,
  \label{eqn:Jpe}
\end{equation}
to the thermal electron collection current, 
\begin{equation}
  J_{\mathrm{e}} = 
e s_{\mathrm{e}} n_{\mathrm{e}} \sqrt{ \frac{k_{\mathrm{B}}
  T_{\mathrm{gas}}}{2\pi m_{\mathrm{e}}}} \left(1+\frac{e\phi}{k_{\mathrm{B}}
  T_{\mathrm{gas}}}\right). \label{eqn:Je}
\end{equation}
Here $s_{\mathrm{e}}
\sim 0.5$ is the electron sticking coefficient, $n_{\mathrm{e}}$ 
is the electron density, obtained in our code from the ionization balance
computation. $T_{\mathrm{gas}}$ is the gas temperature, $m_{\mathrm{e}}$ the electron mass and
$k_{\mathrm{B}}$ the Stefan-Boltzman constant. Eq.~[\ref{eqn:Je}] holds for
positively charged grains ($e\phi > 0$), which is the case for debris
disk gas densities \citepalias{bes07}. Also, as argued by \citet{fer06}, the
charging potential depends logarithmically on disk parameters like
electron density, gas temperature, etc.

$\Gamma_{\mathrm{PE}}$ depends directly on the area covered by the dust
and through eq.~[\ref{eqn:Je}] on electron density $n_{\mathrm{e}}$. If we take
the simplification that each liberated electron carries of order 1\,eV
of energy to the gas (second integral in eq.~[\ref{eqn:Gampe}]), we can
estimate $\Gamma_{\mathrm{pe}} \approx {\mathrm{dust\,area}}\times J_{\mathrm{e}} \times 
1\,\mathrm{eV}/e \propto n_{\mathrm{e}}$ (also see eq. [\ref{eq:PEguess}]).

Also arising from eq.~[\ref{eqn:Je}] is the dependence of PE heating
on gas temperature, $T_{\mathrm{gas}}$. Since $e\phi$ is generally a
few $k_{\mathrm{B}} T_{\star} \gg k_{\mathrm{B}} T_{\mathrm{gas}}$
(\citetalias{bes07}), $e\phi / k_{\mathrm{B}} T_{\mathrm{gas}}$ is the
dominant term in the equation. Hence, $J_{\mathrm{e}}$ (and in turn
$\Gamma_{\mathrm{PE}}$) $\propto T_{\mathrm{gas}}^{-1/2}$. At higher
temperatures, the Coulomb focussing is weakened and electrons are
  less likely to be collected by the positively charged grains, hence
 $e\phi$ becomes more positive, thus raising the $W+e\phi$
threshold. As a result, the PE heating rate is reduced at higher gas
temperatures. There is no runaway heating.

In practice, the double integrals of eq.~[\ref{eqn:Gampe}] 
are pre-computed for a range of $\phi$ values.
As the integrals are monotonic functions of $\phi$, the value of $\phi$
is then obtained by equating the pre-computed thermal
equilibrium calculation $J_{\mathrm{PE}}$ to $J_{\mathrm{e}}$.
\paragraph{Gas-grain collisions.}
The volumetric heating rate due to GG is given by eq.~[19] in
\citetalias{kam01} 
\begin{eqnarray}
\Gamma_{\mathrm{GG}} = 
4.0 \times 10^{-12}{\mbox{erg\,cm$^{-3}$\,s$^{-1}$}}\,\,\,
n_{\mathrm{tot}} \alpha_T \sqrt{T_{\mathrm{gas}}} \nonumber \\
\times (T_{\mathrm{dust}} - T_{\mathrm{gas}})
  \int_{S_{\mathrm{min}}}^{S_{\mathrm{max}}} \pi s^2 
  \frac{dn_{\mathrm{dust}}}{ds} ds,
\end{eqnarray}
where $\alpha_{\mathrm{T}} $ is the thermal accommodation coefficient,
taken to be $0.3$. For $T_{\mathrm{gas}} > T_{\mathrm{dust}}$,
gas-grain collisions cool the gas.
\paragraph{Heating due to photoionization.}
The heating rate per unit volume due to photoionization is
\begin{equation}
\Gamma_{\mathrm{PI}} = \sum_{i=1}^{30} \sum_{j=1}^3 \Gamma_{\mathrm{E},i} n_{i,j},
\label{eqn:PhotoionHeat}
\end{equation}
where $\Gamma_{\mathrm{E},i}$ is the energy released per unit time,
per unit atom ($i$) during photoionization, the subscript $j$ denotes
the ionization state. The photoionization rate is precomputed for a
given stellar flux and recalled when needed.  We follow atoms H
through Zn and through to their second ionization states (see Appendix
for more details).  
\paragraph{Heating due to Gas-Grain Drifting}
  Following BW07, we ignore drift heating due to differential
  velocity between gas and grains. This is justified in our problem
  where gas and grain are weakly coupled, and is also confirmed by the
  calculations of \citet{gor04}.
\subsection{Cooling}
When computing gas cooling, we include the fine structure lines \ion{O}{1}
63.2, 44.1, 145.5\,$\mu$m, \ion{C}{2} 157.7\,$\mu$m and \ion{Si}{2} 34.8\,$\mu$m, 
the atomic cooling lines \ion{O}{1} $\lambda 6\,300$\,\AA\ 
and Ly\,$\alpha$, as well as free-free, and radiative recombination cooling. 

For gas temperatures below 5\,000\,K, \ion{O}{1}, \ion{C}{2} and
\ion{Si}{2} fine-structure transitions dominate the cooling
\citep{hol89}. Collisionally excited levels decay spontaneously,
converting thermal energy to photons that are lost to space,
effectively cooling the gas. While both collisions with free electrons
and atoms/ions can lead to fine-structure cooling, we neglect the
latter since electrons move faster.
\paragraph{Cooling by fine-structure transitions.}
Let $n_i$ be the population density in level $i$.
  For sufficiently high electron densities (and therefore collision
  rates), each level is populated according to the local thermal
  equilibrium (LTE),
\begin{equation} 
n_i = \frac{g_i\exp(-E_{i}/k_{\mathrm{B}} T)} {\sum_i g_i\exp(-E_{i}/k_\mathrm{B} T)} \\\ n_{\mathrm{X}},
\label{eq:ninx}
\end{equation}
where $g_i$ is the statistical weight of state $i$, $E_{i}$ its
  excitation energy, $n_{\mathrm{X}}$ the total number density of
  species X, and the summation runs through all energy levels.  At
  lower densities, a full statistical equilibrium calculation is
  required to determine $n_i$. The threshold densities ($n_{\mathrm{e,
    crit}}$) between SE and LTE are listed in \citet{hol89} for various
  transitions. We collect values for lines of interest to this study
  in Table~\ref{table:critical}.
The typical electron density encountered in our systems is of
  order twenty or less. As such, we have constructed the
  \textsc{ontario} code to perform SE calculation for all elements of
  interest. In particular, for the important cooling elements
  (\ion{O}{1}, \ion{C}{2} and \ion{Si}{2}), the SE calculation is
  self-consistently coupled to the gas thermal equilibrium
  calculation.  Data required in such calculations, as well as the
  numbers of energy levels and transitions included, are detailed in
  the Appendix.

\begin{table}[ht]
\caption{Relevant cooling lines}
\centering
\begin{tabular}{l c c}
\hline
\hline
Line&$A_{ij}$&$n_{\mathrm{e,crit}}$\\
&[s$^{-1}$]&[cm$^{-3}$]\\
\hline
\ion{C}{2} 157.7\,$\mu$m&$2.3\times 10^{-6}$&$8.7 \left( T/100\,\mathrm{K}\right)^{0.50} $\\
\ion{O}{1} 44.1\,$\mu$m&$8.9\times 10^{-10}$&$6.3\times 10^3 \left(T/100\,\mathrm{K}\right)^{-0.03} $\\
\ion{O}{1} 63.2 $\mu$m&$8.9\times 10^{-5}$&$6.3\times 10^3 \left(T/100\,\mathrm{K}\right)^{-0.03} $\\
\ion{O}{1} 145.5\,$\mu$m&$8.9\times 10^{-10}$&$8.9\times 10^2$\\
\ion{Si}{2} 34.8\,$\mu$m&$2.1\times 10^{-4}$&$1.2\times 10^2 \left(T/100\,\mathrm{K}\right)^{0.50} $\\
\hline
\end{tabular}
\tablecomments{
$n_{e,\mathrm{crit}}$ is the critical electron density above which LTE is
approximately correct. $T$ is gas temperature.
}
\label{table:critical}  
\end{table}

The line luminosity per unit volume for fine-structure
  de-excitation $i\rightarrow j$ is contributed by both spontaneous
  and stimulated emission,
\begin{equation}
  F_{10} = [A_{ij}+B_{ij}(U_{\nu)}]\,n_i\:h \nu_{ij},
  \label{eqn:SECoolFlux}
\end{equation}
where $A_{ij}$ and $B_{ij}$ are the usual Einstein
coefficients, $U_{\nu}$ is the radiation field density at
$\nu_{ij}$. We ignore stimulated emission when computing line
fluxes.\footnote{Photons arising from stimulated emission are
  highly anisotropic. In the case of stimulation by stellar photons,
  only emission from gas directly in our line-of-sight to the central
  star is detectable. Moreover, this contribution is negligible
  compared to spontaneous emission from the whole disk. We therefore
  decide to report only line fluxes from spontaneous emission.}
However, we do include the stellar fluxes and dust infrared
  radiation when computing SE.  \citetalias{kam01} demonstrated that
  dust IR radiation promote population in the higher excitation states
  of \ion{O}{1}, leading to stronger \ion{O}{1} line fluxes. 
\edII{The local IR field from a given dust distribution is estimated for each 
position $\mathbf{r}=(\rho,z)$ in the disk by computing the contribution from
100 spherical shells of exponentially increasing radius between 1 and 2000\,AU,
centred on the position $\mathbf{r}$. Each shell is divided
into small sectors (120 in azimuth and 60 in altitude), where the temperatures
and dust densities at the sector positions determine their IR field contribution. 
The contributions from each sector in each shell are then summed up to give the 
total IR field at $\mathbf{r}$.}

The energy absorbed from radiation pumping leaves the system in
  two ways. Part of it is subsequently released as photons by
  spontaneous emission, part of it heats the gas via collisional
  de-excitation. The former process increases the line luminosity (as
  calculated in eq. [\ref{eqn:SECoolFlux}]), but does not contribute
  to the net cooling of the gas. As a result, we define the
  gas cooling rate differently from the cooling line luminosity
  (eq. [\ref{eqn:SECoolFlux}]) as the difference between collisional
  excitation and de-excitation between levels $i$ and $j$,
\begin{equation}
  \Lambda_{ij} = ( q_{ji}\:n_j - q_{ij}\:n_i )\: n_{\mathrm{e}}\:h \nu_{ij},
  \label{eqn:SECoolRate}
\end{equation}
where $q_{ji}$ and $q_{ij}$ are collisional excitation and de-excitation rates, respectively. 
  For very strong radiation fields in the vicinity of early-type stars,
  fluorescence can sometimes cause population inversion and lead to
  effective collisional \emph{heating} of the gas.
\paragraph{\ion{O}{1} (6\,300\,\AA) and Ly\,$\alpha$ transitions.}
Electronic transitions only become important at high
temperatures (eqs.~[34] \& [35] of \citetalias{kam01}):
\begin{eqnarray}
\Lambda_{\mathrm{O\,I, \lambda 6300}}
&=& 1.8 \times 10^{-24} {\mbox{erg\,cm$^{-3}$\,s$^{-1}$}}
n_{\mathrm{O\,I}} n_e\,\,\,
\exp\left(-\frac{22\,800\,\mbox{K}}{T}\right)  \\
\Lambda_{\mathrm{Ly}\,\alpha}
&=& 7.3 \times 10^{-19} {\mbox{erg\,cm$^{-3}$\,s$^{-1}$}}\,\,n_{\mathrm{H\,I}} n_e
\exp\left(-\frac{118\,400\,\mbox{K}}{T}\right).
\end{eqnarray}
\paragraph{Free-free cooling.}
Free-free cooling also enters at high temperatures. Cooling due
to electrons interacting with ions of charge Z is \citep[eq.~{[3.14]} in][]{ost89}
\begin{equation} 
\Lambda_{\mathrm{ff}}(Z)
1.42 \times 10^{-27} {\mbox{erg\,cm$^{-3}$\,s$^{-1}$}} \,\,
g_{\mathrm{ff}}\,Z^2 \sqrt{T} n_e n_{\mathrm{+}},
\end{equation}
where $n_+$ is the ion number density and
$g_{\mathrm{ff}}
\approx 1.3$ is the mean Gaunt factor for free-free emission. The
total free-free cooling rate is the summation over all ionic
species. The recombination cooling is similar to that in
eq.~[\ref{eqn:PhotoionHeat}], but with $\Gamma_{\mathrm{E}}$ replaced by
$\Lambda_{\mathrm{E}}$, the recombination cooling rate per electron for
species $i$, and with $n_{i,j}$ replaced by electron number density
$n_{\mathrm{e}}$:
\begin{equation}
\Lambda_{\mathrm{E}} = \sum_{i=1}^{30} \sum_{j=1}^3 \Lambda_{\mathrm{E},i} n_{\mathrm{e}}.
\label{eqn:Recomb}
\end{equation}

\section{The $\beta$~Pictoris debris disk}
\label{sec:BPic}
In order to check the reliability of \textsc{ontario},
we apply it to the well-studied debris disk of $\beta$~Pictoris.
The $\beta$\,Pic system is unique in that it is the only debris system
where both the dust and the gas components have been extensively mapped
\citep[see, e.g., ][and references therein]{che07}.

\subsection{Input parameters}
\label{subsec:bpic-input}

The adopted model parameters of the $\beta$\,Pic system are listed in
Table~\ref{table:Params_BPic}. We used the same flux-calibrated
stellar model spectrum for $\beta$\,Pic as in \citet{fer06}, except we
merged the model spectrum with far- and near-ultraviolet observations
from \textit{FUSE} and \textit{HST/STIS} \citetext{A.\ Roberge, priv.\
  comm.}, covering 925--1\,180\,\AA\ and 1\,465--1\,660\,\AA,
respectively, and interpolating over the gap.  The distribution of
dust area is deduced from \textit{HST/STIS} observations
\citep{hea00,fer06} and  summarized by a fitting form
\begin{equation}
  \pi \langle a^2 \rangle n_{\mathrm{dust}} = \frac{\tau_0}{W}
  \frac{\exp\left[-(z/W )^{0.7}\right]}
  {\sqrt{(\rho/\rho_0)^{-4}+(\rho/\rho_0)^6}},
  \label{eqn:n_dust_BPic}
\end{equation}
where $\rho$ is the cylindrical radius and $z$ the height above
disk midplane. The empirically determined fitting parameters are
$\rho_0 = 120$\,AU, $W = 6.6(\rho/\rho_0)^{0.75}$\,AU, and $\tau_0 = 2
\times 10^{-3}$. We also adopt the following form to describe the
  distribution of the total gas number density \citep{bra04}:
\begin{equation}
  n_{\mathrm{gas}}(\rho,z) = n_0\left[\frac{2}{(\rho/\rho_1)^{2\alpha_1} + (\rho/\rho_1)^{2\alpha_2}}\right]^{1/2}
  \,\,\exp\left[-\frac{z^2}{2\sigma_{\mathrm{z}}^2}\right].
  \label{eqn:n_gas_BPic}
\end{equation}
For elemental abundances, we assume solar abundances \citep{gre93}
  except for three elements: hydrogen is set to $10^{-3}$ of its solar
  value, helium is assumed to be zero, and carbon is set to be
  $20\times$ its solar abundance. This is motivated below. We then
  compare our model output to the observed \ion{Na}{1} profile to
  obtain values for the fitting parameters, $\rho_1 \approx 100$\,AU,
  $\alpha_1 = 1.0$, $\alpha_2 = 2.3$, $n_0 = 24$\,cm$^{-3}$, $\sigma_z
  = (\rho / 100\,\mathrm{AU}) \times 40\,\mathrm{AU} / \sqrt{8\ln
    8}$.
This derived distribution is similar to that in eq.~[4] of
  \citet{bra04} except for two differences: our $n_0$ is a factor of
  $10^3$ smaller as we adopt a hydrogen poor mix while they assumed a
  solar abundance of hydrogen; our vertical distribution is well
  approximated by a Gaussian profile with a full width half maximum
  (FWHM) of $H = 0.4 r$. This differs from \cite{bra04} because we
  include the variation of ionization fraction with height in this
  work, while \citet{bra04} did not.

The volume density of dust area (eq.~[\ref{eqn:n_dust_BPic}]) vanishes
close to the star, indicating an inner clearing of dust. The gas
profile of the inner disk is uncertain. The \ion{Na}{1} emission line
can be traced in to a projected separation of 13\,AU from the star and
out to 323\,AU from the star.  Comparing the \ion{Na}{1} seen in
emission with the \ion{Na}{1} seen in absorption, \citet{bra04}
concludes that 80--100\,\% of the circumstellar \ion{Na}{1} seen in
absorption is within this range.  We therefore limit the gas disk to
start at 15\,AU and end at 200\,AU (with more distant regions making
negligible contributions to the total line luminosities). The model
disk height is limited to 100\,AU above the midplane.

To relate the dust opacity distribution of eq.~[\ref{eqn:n_dust_BPic}]
to the dust disk mass, we assume that dust grains are spherical and
range in sizes $s_{\mathrm{min}} < s < s_{\mathrm{max}}$, with a
differential number distribution $dn_{\mathrm{dust}}/ds \propto s^{-3.5}$,
as characteristic of dust in the interstellar
medium. $s_{\mathrm{max}}$ is set, somewhat arbitrarily, to 1\,cm,
while $s_{\mathrm{min}}$ is set to be the minimum size below which
radiation pressure could remove the grains. For a grain initially at
circular orbit, this occurs when the ratio of the radiation pressure
to gravitational force acting on the grains, $\beta$, is equal to 0.5.
This translates to
\begin{equation}
  s_{\mathrm{min}} = \frac{3 L_{\star} Q_{\mathrm{PR}} }{16 \pi G M_{\star} c \varrho_{\mathrm{grain}} \beta}
= \frac{3 L_{\star} Q_{\mathrm{PR}} }{8 \pi G M_{\star} c \varrho_{\mathrm{grain}}},
\label{eq:smin}
\end{equation}
where $c$ is the speed of light and $Q_{\mathrm{PR}} \sim 1$ is
  the radiation pressure efficiency averaged over the stellar
  spectrum. For $\beta$\,Pic, we assume $M_{\star}$ =
1.75\,$M_{\odot}$ and $L_{\star}$ = 11\,$L_{\odot}$. Taking
$\varrho_{\mathrm{grain}} \sim 1$\,g\,cm$^{-3}$, we find
$s_{\mathrm{min}} \sim$ 5\,$\mu$m.  

The infrared luminosity from the dust, integrated over the whole
 disk, is thus given by:
\begin{equation}
  L_{\mathrm{dust}} = \int \hspace{-.4em} \int \frac{L_{\star}}{4 \pi (\rho^2 + z^2)} (1-\epsilon)
\pi \langle a^2 \rangle n_{\mathrm{dust}}
2 \pi \rho \,\, d\rho dz,
\end{equation}
where $\epsilon \sim 0.5$ is the grain reflectivity coefficient.
When using the $\beta$\,Pic dust distribution described by
eq.~[\ref{eqn:n_dust_BPic}], $L_{\mathrm{dust}} \sim 10^{-3}
L_{\star}$.

To derive the grain temperature as a function of distance from
the star, we use an empirical relation derived from observations
of the $\beta$\,Pic dust disk. From observations (e.g.\ Nilsson et 
al.\ 2009) it is clear that the dust cannot radiate as black bodies,
as that would imply too much sub-mm radiation. Instead, the spectral 
energy distribution of the disk is found to be well fit by a modified
black body,
\begin{equation}
\label{eq:grey}
I_{\nu} = \left(\frac{\nu}{\nu_0}\right)^{\gamma} B_{\nu}(T).
\end{equation}
One can
attempt to relate $\gamma$ and $\nu_0$ to dust properties \citep[e.g.][]{dra03},
but that is beyond the scope of the present article, where we are content
that this simple empirical relation approximates the observed dust emission 
surprisingly well. Using eq.~[\ref{eq:grey}] for radiative equilibrium,
where we equate the absorbed energy flux to the emitted, we get
\begin{equation}
\label{eq:dusttemp0}
\frac{(1-\epsilon)L_{\star}}{4\pi R^2} \pi a^2 = 4\pi a^2 \int_0^{\infty}\pi\left(\frac{\nu}{\nu_0}\right)^{\gamma} B_{\nu}(T),
\end{equation}
which in turn implies
\begin{equation}
\label{eq:dusttemp1}
T = \frac{(1-\epsilon)L_{\star}\nu_0^2c^2}{32h\pi^2R^2\zeta(4+\gamma)\Gamma(4+\gamma)},
\end{equation}
where $\epsilon$ is an average albedo for stellar radiation, $L_{\star}$ is
the luminosity of the star, $h$ is Planck's constant, $c$ the speed of light,
$\zeta$ is the Riemann $\zeta$-function, and $\Gamma$ is the (true) gamma
function. Using eq.~[\ref{eq:dusttemp1}] to compute the temperatures the
$\beta$\,Pic dust disk, and eq.~[\ref{eq:grey}] and the dust area density
distribution of eq.~[\ref{eqn:n_dust_BPic}] to compute the emission, we integrate the
total emission from the disk seen at Earth to fit $\gamma = 0.67$ and 
$\nu_0 = 4\times10^{14}$\,Hz by comparing to the SED summarized in 
\citet{nil09}. A simplified expression for the dust temperature is
thus
\begin{equation}
\label{eq:dusttemp2}
T \approx 430\,
\mathrm{K}\,\left(\frac{L_{\star}}{L_{\odot}}\right)^{0.214}
\left(\frac{R}{\mathrm{AU}}\right)^{-0.428}
\left(\frac{\nu_0}{4\times10^{14}\,\mathrm{Hz}}\right)^{0.143}.
\end{equation}

Returning back to the issue of elemental abundances, observations
and numerical modeling of gas in the $\beta$\,Pic system indicate that
the disk may be deficient in hydrogen, consistent with a disk
comprised of mainly metals, produced as a result of dust sublimation
\citep{lec01,bra04,fer06}. In addition, the gas in the disk has been
found to be of 20$\times$ higher carbon abundance than expected for
solar composition \citep{rob06}. Other observed elements are close to
solar in relative abundance. The abundance of oxygen, which is
important for cooling the gas, is unfortunately hard to constrain,
because the available absorption lines are strongly saturated
\citep{rob06}, although a solar abundance is consistent with
data. Consequently, for our study, we set all elemental abundances at
their solar values except for carbon at 20 times solar, hydrogen at
$10^{-3}$ solar and helium abundance set to zero.

\begin{table}[ht]
\caption{$\beta$\,Pic disk model input parameters}
\centering
\begin{tabular}{l l}
\hline
\hline
M$_\star$&1.75\,$M_\odot$ \\
R$_\star$&2\,$R_\odot$ \\
L$_\star$&11\,$L_\odot$ \\
Disk inner cutoff & 15\,AU \\
Dust mass & 0.27\,$M_{\oplus}$\\
Dust profile & eq.~[\ref{eqn:n_dust_BPic}] \\
Dust luminosity& $9.6 \times 10^{-4}$\,$L_{\star}$\\
Gas mass & $8.3 \times 10^{-4}$\,$M_{\oplus}$ \edII{(assuming H-depleted gas)}\\
Gas profile & eq.~[\ref{eqn:n_gas_BPic}] \\
Elemental abundances & solar \citep{gre93}, except \\
&  [C] = 20\,[C]$_{\odot}$, [H] = $10^{-3}$\,[H]$_{\odot}$ and [He] = 0 \\
\hline
\end{tabular}
\label{table:Params_BPic}
\end{table}

\subsection{Comparing the model to observations}
\label{subsec:bpic-output}

\subsubsection{Disk temperature profile}
\label{sec:BPic_Temp}

Fig.~\ref{fig:bPic_vs_rho} shows $\beta$\,Pic disk midplane gas and
dust distribution and temperature profiles together with important
heating and cooling mechanisms. The temperature in the disk is
determined by two major heating mechanisms: PE and PI heating. PE is
maximum around 100 AU where the dust distribution peaks; PI dominates
in the strong radiation field within $\sim\,25$ AU of the star. With
low gas densities gas-grain collision rates are negligible, so that
gas and dust temperatures are effectively decoupled. The result is a
gas temperature profile that peaks at the inner disk boundary and at
100 AU. The cooling is dominated almost exclusively by the \ion{C}{2}
157.7\,$\mu$m fine structure transition. The \ion{Si}{2} line transition
becomes stronger in the higher temperature region around dust
distribution peak, yet still remains $\sim$ 100 times below the
\ion{C}{2} flux.

\subsubsection{Column densities}

\begin{figure}
\begin{center}
   \includegraphics[width=\columnwidth]{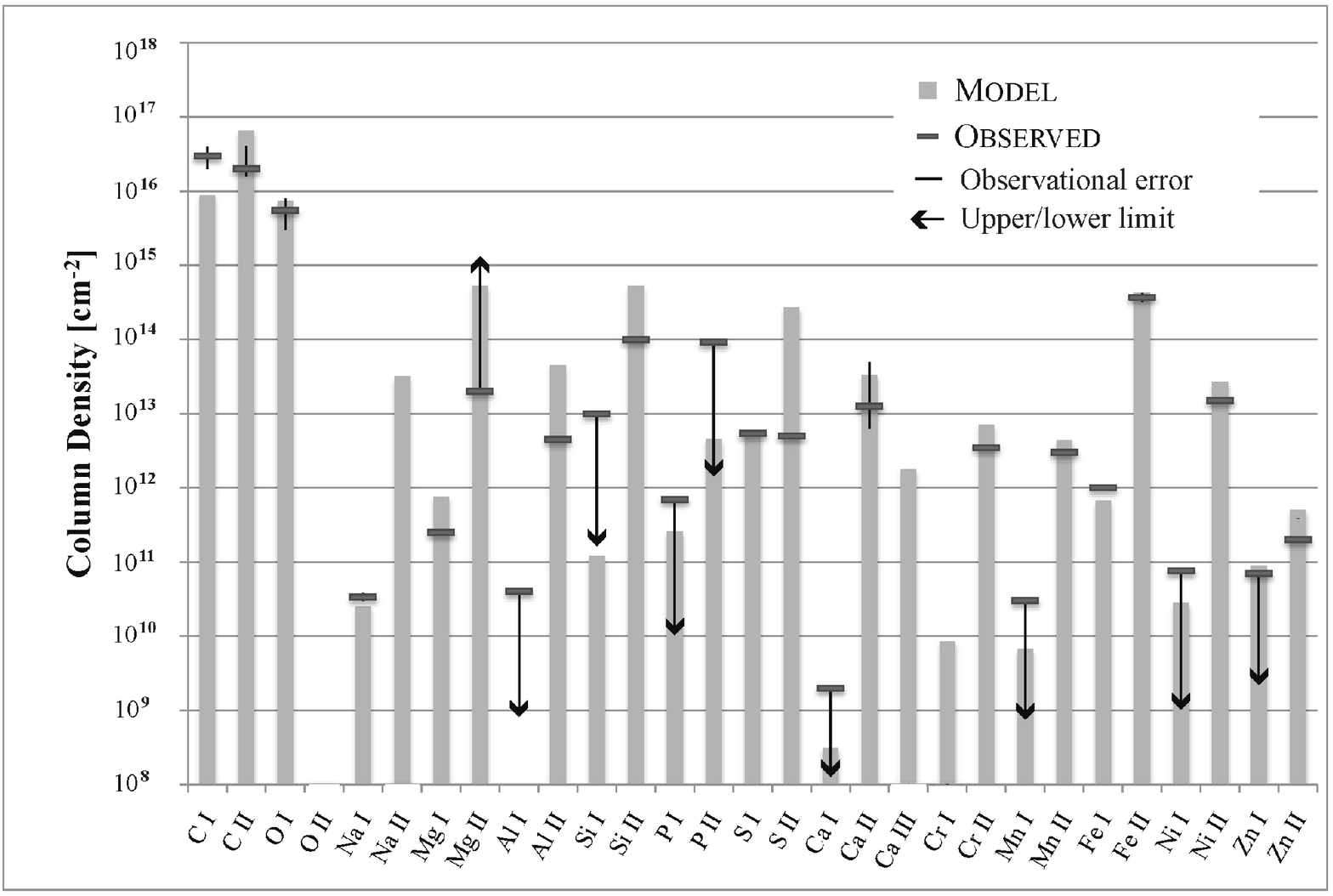} 
   \caption{Column densities for metallic gas in the $\beta$\,Pic
     disk, same data as in Tab.~\ref{table:ColDens}. No data for \ion{O}{2}, \ion{Na}{2}, \ion{Ca}{3},
     \ion{Cr}{1} are available as these are not observed in optical spectra. \ion{Na}{1} and \ion{Fe}{2} are strongly constrained, and no observational uncertainty 
     data are available for \ion{Mg}{1}, \ion{Al}{2}, \ion{Si}{2}, \ion{S}{1}, \ion{S}{2},
     \ion{Cr}{2}, \ion{Mn}{2}, \ion{Fe}{1}, \ion{Ni}{2}, \ion{Zn}{2}.}
\label{fig:ColumnDens}
\end{center}
\end{figure}

\begin{figure*}
\begin{center}
   \includegraphics[width=\textwidth]{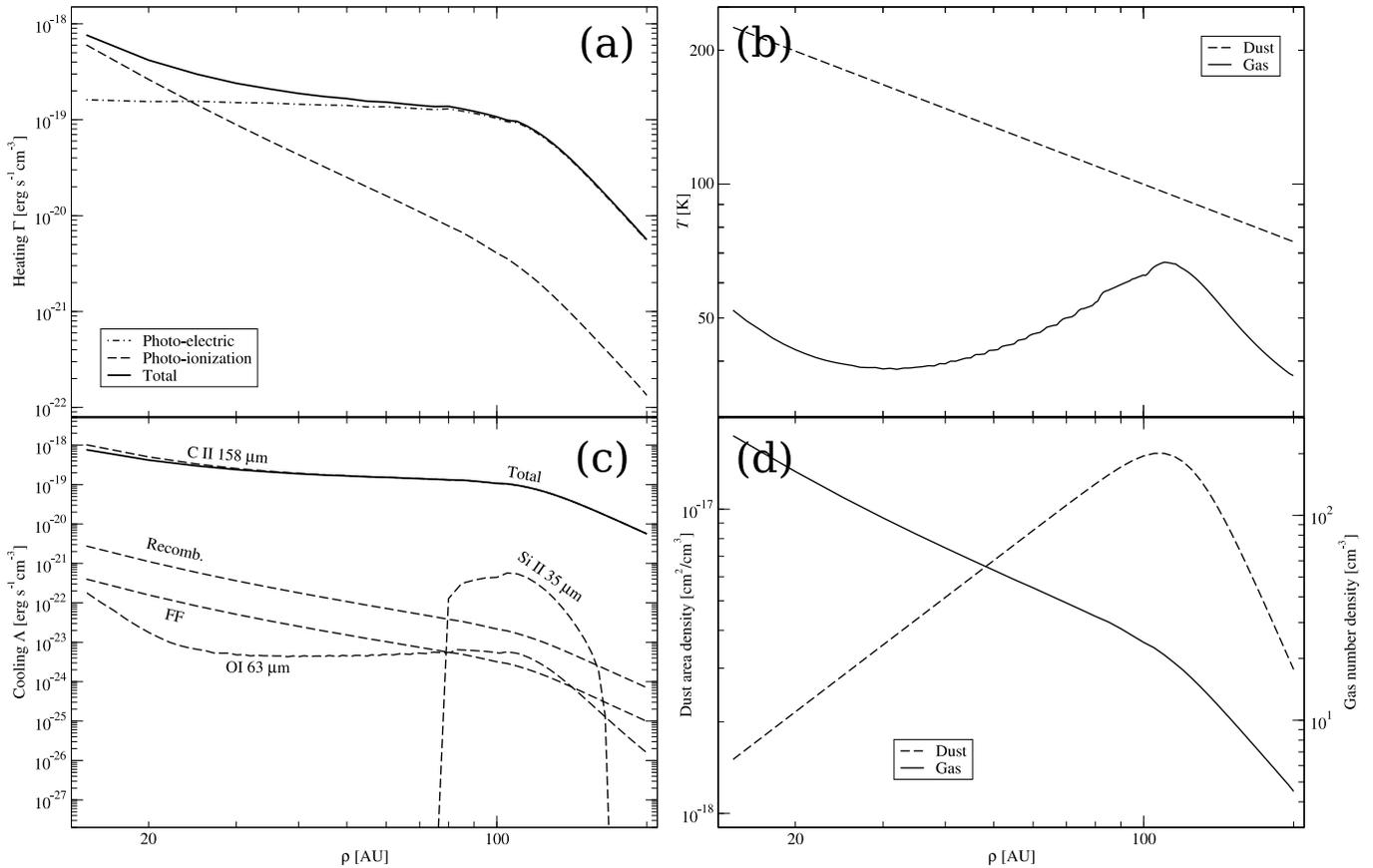} 
\caption{ Physical
   properties of the $\beta$\,Pic midplane in our model. The upper
   left panel (a) shows the few important heating rates, and the lower
   left panel (c) the dominant cooling rates. The upper right panel (b) shows
   the gas and dust temperatures, respectively, and the lower right
   panel (d) gas and dust densities (note different scales and units for
   the two populations).}  \label{fig:bPic_vs_rho}
\end{center}
\end{figure*}

The edge-on orientation of the $\beta$\,Pic disk makes it possible to
observe the disk gas through absorption. In Table~\ref{table:ColDens}
we list column densities inferred from observations and column
densities computed with \textsc{ontario}, using the disk model
outlined in \S\,\ref{subsec:bpic-input}.  The same comparison is
  also presented in Fig.~\ref{fig:ColumnDens} for better
  visualization.
\begin{table}[ht] 
\caption{Column densities for metallic gas in the $\beta$\,Pic disk}
\centering
\begin{tabular}{l r r}
\hline
Species & Observed & Model \\ & [cm$^{-2}$] & [cm$^{-2}$]\\
\hline
\ion{C}{1}& (2--4)$\times 10^{16}$ & $8.9\times 10^{15}$ \\
\ion{C}{2}& $2.0^{+2.1}_{-0.4}\times 10^{16}$ & $6.6\times 10^{16}$ \\
\ion{O}{1}& (3--8)$\times 10^{15}$ & $7.4\times 10^{15}$ \\
\ion{O}{2}& \nodata & $0.0$ \\
\ion{Na}{1}& $(3.4\pm 0.4)\times 10^{10}$ & $2.5\times 10^{10}$ \\
\ion{Na}{2}& \nodata & $3.2\times 10^{13}$ \\
\ion{Mg}{1}& $2.5\times 10^{11}$ & $7.6\times 10^{11}$ \\
\ion{Mg}{2}& $\geq 2\times 10^{13}$ & $5.3\times 10^{14}$ \\
\ion{Al}{1}& $\leq 4\times 10^{10}$ & $1.0\times 10^{8}$ \\
\ion{Al}{2}& $4.5\times 10^{12}$ & $4.5\times 10^{13}$ \\
\ion{Si}{1}& $\leq 1\times 10^{13}$ & $1.2\times 10^{11}$ \\
\ion{Si}{2}& $1\times 10^{14}$ & $5.3\times 10^{14}$ \\
\ion{P}{1}& $\leq 7.0\times 10^{11}$ & $2.6\times 10^{11}$ \\
\ion{P}{2}& $\leq 9.2\times 10^{13}$ & $4.6\times 10^{12}$ \\
\ion{S}{1}& $5.4\times 10^{12}$ & $6.1\times 10^{12}$ \\
\ion{S}{2}& $5\times 10^{12}$ & $2.7\times 10^{14}$ \\
\ion{Ca}{1}& $\leq 2\times 10^9$ & $3.1\times 10^8$ \\
\ion{Ca}{2}& $1.26^{+3.75}_{-0.63}\times 10^{13}$ & $3.3\times 10^{13}$ \\
\ion{Ca}{3}& \nodata & $1.8\times 10^{12}$ \\
\ion{Cr}{1}& \nodata & $8.5\times 10^{9}$ \\
\ion{Cr}{2}& $3.5\times 10^{12}$ & $7.1\times 10^{12}$ \\
\ion{Mn}{1}& $\leq 3\times 10^{10}$ & $6.7\times 10^9$ \\
\ion{Mn}{2}& $3\times 10^{12}$ & $4.4\times 10^{12}$ \\
\ion{Fe}{1}& $1\times 10^{12}$ & $6.7\times 10^{11}$ \\
\ion{Fe}{2}& $(3.7\pm 0.5)\times 10^{14}$ & $4.3\times 10^{14}$ \\
\ion{Ni}{1}& $\leq 7.6\times 10^{10}$ & $2.8\times 10^{10}$ \\
\ion{Ni}{2}& $1.5\times 10^{13}$ & $2.7\times 10^{13}$ \\
\ion{Zn}{1}& $\leq 7\times 10^{10}$ & $8.9\times 10^{10}$ \\
\ion{Zn}{2}& $2\times 10^{11}$ & $5.1\times 10^{11}$ \\
\hline
\end{tabular}
\tablecomments{Observed column densities are compiled by \citet{rob06}. Some
species are unconstrained, as no absorption line has been
observed. The column density of \ion{Mg}{2} is to be considered a lower limit
since the line is strongly saturated.}
\label{table:ColDens}  
\end{table}

The real gas disk is far from being cylindrically symmetric, with the
observed
\ion{Na}{1} profiles from the south-west (SW) and north-east (NE) parts of the
$\beta$\,Pic disk differing substantially. The NE side is the
brightest within a projected distance of 35\,AU and outside 100\,AU,
and is detected to the limit of the observations at 323\,AU. The SW
side on the other hand, is slightly brighter in the 35--100\,AU
region, and drops much more quickly outside 100\,AU than the NE side
\citep[Fig.~3 of ][]{bra04}.
Given that the model is based on the spatially resolved emission from the 
\ion{Na}{1} D$_2$ line, where the \ion{Na}{1} represents only
10$^{-3}$ of the total Na, the agreement in column density between
observation and prediction is remarkable, and reinforces the conclusion by 
\citet{bra04} that the gas seen in absorption is the same seen in emission. 

The model predicts most elements to be highly ionized, with the
exceptions of O and C, due to their higher ionization potential. Only
a few elements have their ionization fractions directly determined (as
an column density average), but many have ionization fractions
constrained by the data. Overall, the predicted ionization levels from
\textsc{ontario} are consistent with observations, with the exception
of C and S, which appear less ionized than predicted. Interestingly,
together with the elements O and P (which do not have their ionization
level observationally constrained), these are the elements of
Table~\ref{table:ColDens} with the highest ionization potential (first
ionization potentials are 10.4 [for S], 11.3 [C], 13.6 [O], and
10.5\,eV [P]).  Their ionization fractions are thus the most sensitive
to the stellar UV spectrum at energies higher than 10\,eV. A
possibility for the discrepancy is thus that the UV flux in our
adopted stellar spectrum is overestimated compared to the actual
spectrum.

There is the tantalizing possibility of doing cosmo-chemistry using
these data. If indeed the metallic gas arises from grain-grain
collisions, the evaporated gas could have the same chemical
composition as the grains. This provides a window to study the make-up
of $\beta$\,Pic's planetesimal belt, much like what cosmochemists have
accomplished by studying meteorites that fall on Earth. Among the
intriguing questions to ask are: why is C so super-solar, 
and is the factor of $10$ depletion observed in Al significant?

\subsubsection{Cooling line fluxes}

As outlined in \S\,\ref{sec:heating}, we can estimate the total
cooling line fluxes by estimating the PE heating rate.  Assuming that
every escaping electron heats the gas by $\sim$1\,eV, we obtain the
following order-of-magnitude estimate for the cooling luminosity,
\begin{eqnarray}
L_{\mathrm{cool}} & \sim & \int dV \Lambda_{\mathrm{fine}}  \approx 
\int dV \Gamma_{\mathrm{pe}} \nonumber \\
& \approx   &
\int dV \int_{s_{\mathrm{min}}}^{s_{\mathrm{max}}} ds {\frac{dn_{\mathrm{dust}}}{ds}} \pi s^2 \times 
\frac{J_{\mathrm{e}}}{e} \times 1\,\mathrm{eV} \nonumber\\ 
& \sim & {\frac{2L_{\mathrm{dust}}}{L_*}} \times 4\pi R^2 \times {\frac{J_{\mathrm{e}}}{e}}
\times 1\,\mathrm{eV} \nonumber \\ & \sim &  10^{-8} L_{\odot}
\left({\frac{T_{\mathrm{gas}}}{100\,\mathrm{K}}}\right)^{-1/2} \left({\frac{R}{100\,\mathrm{AU}}}\right)^2\nonumber \\
& & \times \left({\frac{L_{\mathrm{dust}}/L_*}{10^{-3}}}\right)\,
\left({\frac{n_{\mathrm{e}}}{20\,\mathrm{cm}^{-3}}}\right)\, \left({\frac{e \phi}{2\,\mathrm{eV}}}\right),
\label{eq:PEguess}
\end{eqnarray}
where the factor of $2$ in front of $L_{\mathrm{dust}}$ corrects for
albedo, and $R$ is the radius of the dust ring.  This estimate
roughly explains the total cooling luminosity of $\sim 7\times
  10^{-8} L_\odot$ we obtain for $\beta$\,Pic (Table~\ref{table:CoolFlux_BPic}), 
and it illuminates the dependency of
cooling luminosity on gas and dust properties.

We compare these values to a previous model by \citetalias{bes07},
also shown in Table \ref{table:CoolFlux_BPic}. The total cooling
luminosity there is higher by a factor of 30 and the dominant line is
\ion{O}{1} 63.2$\mu$m line, as opposed to the \ion{C}{2} 157.7$\mu$m
line here. These are explained by the three major differences between
the model we adopt here and that adopted by \citetalias{bes07}: 1) the
gas mass (the metallic component) adopted in \citetalias{bes07} is
about seven times greater than we use here. As they also include
contribution to electron density from hydrogen, the total electron
density is some fifteen times greater than our value (see
eq.~[\ref{eq:PEguess}]). 2) \citetalias{bes07} used LTE to compute the
level occupation, while we calculate the detailed statistical
equilibrium (SE).  For the low electron density in the $\beta$\,Pic
disk ($n_{\mathrm{e}} \sim 20$\,cm$^{-3} \ll n_{\mathrm{e,crit}}$ for
some ions), an LTE treatment grossly overestimate the occupation
number in the excited states of \ion{O}{1}. This accounts for their
strong \ion{O}{1} lines. In our $\beta$\,Pic model, oxygen atoms do no
contribute to cooling significantly.  
\begin{table}[ht] 
\caption{$\beta$\,Pic cooling line luminosities}
\centering
\begin{tabular}{l c c c c}
\hline
\hline
Line&\multicolumn{3}{c}{\textsc{ontario} output}& \citetalias{bes07} \\
\hline
& Luminosity & Flux & Flux & Flux\\
& & & (no dust IR) &\\
& [$L_{\odot}$] &[erg\,s$^{-1}$\,cm$^{-2}$]&[erg\,s$^{-1}$\,cm$^{-2}$]&[erg\,s$^{-1}$\,cm$^{-2}$]\\
\hline
\ion{C}{2} 157.7\,$\mu$m& $9.0\times 10^{-8}$  & $7.2\times 10^{-15}$ & $7.2\times 10^{-15}$ &$6.0\times 10^{-15}$\\
\ion{O}{1}  44.1\,$\mu$m& $1.9\times 10^{-15}$ & $1.5\times 10^{-22}$ & $1.5\times 10^{-22}$ &$7.7\times 10^{-20}$\\
\ion{O}{1}  63.2\,$\mu$m& $7.1\times 10^{-10}$ & $5.7\times 10^{-17}$ & $5.5\times 10^{-17}$ &$2.1\times 10^{-13}$\\
\ion{O}{1} 145.5\,$\mu$m& $7.3\times 10^{-11}$ & $5.9\times 10^{-18}$ & $5.9\times 10^{-18}$ &$3.1\times 10^{-15}$\\
\ion{Si}{2} 34.8\,$\mu$m& $5.6\times 10^{-9}$  & $4.5\times 10^{-16}$ & $4.5\times 10^{-16}$ & \nodata \\
\hline
\end{tabular}
\tablecomments{{Luminosities} and fluxes of the important cooling lines,
as observed on Earth, computed by \textsc{ontario}. To estimate the contribution from the dust IR field, 
cooling line fluxes without dust IR field are included.}
\label{table:CoolFlux_BPic}
\end{table}
To test that there are no other significant differences between
\textsc{ontario} and the model used by \citetalias{bes07}, we used
their gas and dust distribution and enforced LTE for the level
populations, and confirmed that these changes made \textsc{ontario}
reproduce the \citetalias{bes07} results.

For the $\beta$\,Pic dust and gas profiles, turning off the dust
  IR field in our code only produces a $<$4\,\% variation on the
  \ion{O}{1} 63.2\,$\mu$m line flux.

Our predicted value for the \ion{C}{2} 157.7\,$\mu$m line flux is high 
enough to be detectable with heterodyne far-infrared spectrometer
\textit{HIFI} on \textit{Herschel}. For a crude estimate of the
instruments sensitivity, we use the pre-launch predicted 
sensitivity, expressed as the system temperature 
$T_{\mathrm{sys}} = 2\,000$\,K at 157.7\,$\mu$m. Assuming the emission to be
unresolved at a nearly diffraction-limited beam-size of 12\arcsec, a flux of
$7.2\times 10^{-15}$\,erg\,s$^{-1}$\,cm$^{-2}$ corresponds to 
the velocity integrated antenna temperature 0.38\,K\,km\,s$^{-1}$.
Assuming the line to be 10\,km\,s$^{-1}$ broad, i.e.\ the frequency 
bandwidth $\Delta\nu = 60$\,MHz, the effective 
integration time required for a $5\sigma$-detection should be
on the order of 
\begin{equation}
t \sim \frac{1}{6\times10^{7}\,\mathrm{Hz}} \left(\frac{5 \times 10\,\mathrm{km}\,\mathrm{s}^{-1} \times 2000\,\mathrm{K}}
{0.38\,\mathrm{K}\,\mathrm{km}\,\mathrm{s}^{-1}}\right)^2 = 20\,\mathrm{min}.
\end{equation}
The line could also be detected by \textit{PACS} on \textit{Herschel}, with
a similar effective integration time.

A tentative detection of the \ion{C}{2} 157.7\,$\mu$m line was reported
for the $\beta$\,Pic disk \citep{kam03} with a flux of
$10^{-13}$\,erg\,cm$^{-2}$\,s$^{-1}$ ($4\sigma$ result). 
This is more than 10$\times$ stronger than our prediction and
a confirmation by Herschel is necessary (we estimate it would be 
confirmed with a $5\sigma$ confidence in $\sim 6$ seconds of integration
time).

Unfortunately, the \ion{O}{1} 63.2\,$\mu$m emission predicted by our
model ($5.7\times 10^{-17}$\,erg\,s$^{-1}$\,cm$^{-2}$) is far too
weak to be detected by \textit{PACS}, which has
a pre-launch estimated sensitivity of $10^{-15}$\,erg\,s$^{-1}$\,cm$^{-2}$
(5$\sigma$, 1\,h). The wavelength region of \ion{Si}{2}
34.8\,$\mu$m line is not covered by any present or planned instrument
and, as such, cannot be confirmed.

\section{LINE LUMINOSITIES FROM DISKS AROUND A AND F STARS}
Having studied the $\beta$\,Pic disk in some detail, we now
  proceed to investigate a range of potential debris disk
  configuration. We first study a fiducial disk with parameters tuned
  to resemble the $\beta$\,Pic disk of \S\,\ref{sec:BPic}, but with
  simplified gas and dust profiles.  We then investigate the
  temperature and ionization rates in the disk, fine-structure cooling
  line fluxes, and identify major emission and absorption lines. We
  vary these parameters systematically and report their
  influences on the observed cooling fluxes. In particular, we sample
  changes in elemental abundances (C, O, and H), variations in the
  total mass and distribution of gas and dust in the disk, and test a
  range of stellar spectral types.
\subsection{Fiducial disk}
\label{sec:Fiducial}
\begin{table}[ht]
\caption{Fiducial Case Input Parameters}
\centering
\begin{tabular}{l l}
\hline
\hline
$\mathrm{M}_{\star}$ & 1.75\,$M_{\odot}$ \\
$\mathrm{R}_{\star}$ & $2\,R_{\odot}$ \\
$\mathrm{L}_{\star}$ & 11\,$L_{\odot}$ \\
Dust mass & 0.5\,$M_{\oplus}$ (assuming a bulk density of \\
          & 1\,g\,cm$^{-3}$ and $s_{\mathrm{max}}=1$\,cm) \\
Dust profile & eq.~[\ref{eqn:n_GasDust_Fid}]: $\rho_{0,\mathrm{dust}} = 120$\,AU, FWHM$_{\rho} = 10$\,AU\\
&$H/\rho$ = 0.1, $n_{0,\mathrm{dust}}$  = $2.9\times10^{-16}$ cm$^2$\,cm$^{-3}$\\
Dust size distribution & $dn_{\mathrm{dust}}/ds \propto s^{-3.5}$, $s_{\mathrm{min}} \le s \le $1\,cm \\
Gas mass & $3.0\times 10^{-4}$\,$M_\oplus$\\
Gas profile & eq.~[\ref{eqn:n_GasDust_Fid}]: $n_{0,\mathrm{gas}} = 24$\,cm$^{-3}$, $\rho_{0,\mathrm{gas}} = 100$\,AU,\\
&FWHM$_{\rho} = 40$\,AU, $H/\rho$ = 0.4 \\
Elemental abundances& solar \citep{gre93}, except \\ 
& [C] = 20\,[C]$_{\odot}$, [H] = $10^{-3}$\,[H]$_{\odot}$ and [He] = 0 \\
\hline
\label{tbl:InputParams_Fid}
\end{tabular}
\end{table}
Fiducial case input parameters are listed in table \ref{tbl:InputParams_Fid}. For the gas
distribution observed in the $\beta$\,Pic system
(eq.~[\ref{eqn:n_gas_BPic}]), the total mass of the gas disk depends
not only on the parameters specifying the gas profile, but also on the
choice of inner and outer disk boundaries. To avoid this complication,
we choose a double-Gaussian profile for both the gas and the dust
components,
\begin{equation}
  n= n_{0} \exp \left[-\frac{(\rho-\rho_{0})^2}{2\sigma_{\rho}^2}
  \right] \,\,
  \exp\left[-\frac{z^2}{2\sigma_{\mathrm{z}}^2}\right].
  \label{eqn:n_GasDust_Fid}
\end{equation}
Here $\rho$ and $z$ are cylindrical coordinates, $\sigma_{\rho} =
\mathrm{FWHM}_{\rho}/\sqrt{8\ln 2}$ represents the width of the radial
distribution and $\sigma_{\mathrm{z}} =
(\rho/\rho_{\mathrm{ref}})\times \mathrm{FWHM}_{\mathrm{z}}/\sqrt{8\ln
2}$ is the vertical profile width that scales linearly with distance.
At $\rho_{\mathrm{ref}} = 100$\,AU we set FWHM$_{\mathrm{z}}$ to
40\,AU, giving a constant $H/\rho$ = 0.4.  The peak midplane density
$n_{0,gas}$ is set to 24\,cm$^{-3}$, based on the value observed for the
$\beta$\,Pic system. The interstellar medium gas density of $6.7\times
10^{-3}$\,cm$^{-3}$ is chosen as a lower limit throughout the disk,
should the Gaussian of eq.~[\ref{eqn:n_GasDust_Fid}] fall below this
($6.7\times 10^{-3}$\,cm$^{-3}$ corresponds to 1\,cm$^{-3}$ if solar
abundances of H and He were to be included).  The Gaussian
distribution of the dust is slightly offset from the gas in the radial
direction (see Fig. \ref{fig:vs_Rho}) with a steeper radial and
vertical drop-off (FWHM$_{\rho}$ of 10\,AU and $H/\rho$ = 0.1).

The implied dust luminosity is $1.2\times 10^{-3}L_{\star}$.  To
convert between the dust area distribution and the total dust mass, we
computed $s_{\mathrm{min}}$ from eq.~[\ref{eq:smin}]. 

\subsubsection{Disk temperature profile}
\label{sec:Fid_Temp}

\begin{figure*}
\begin{center}
\includegraphics[width=\textwidth]{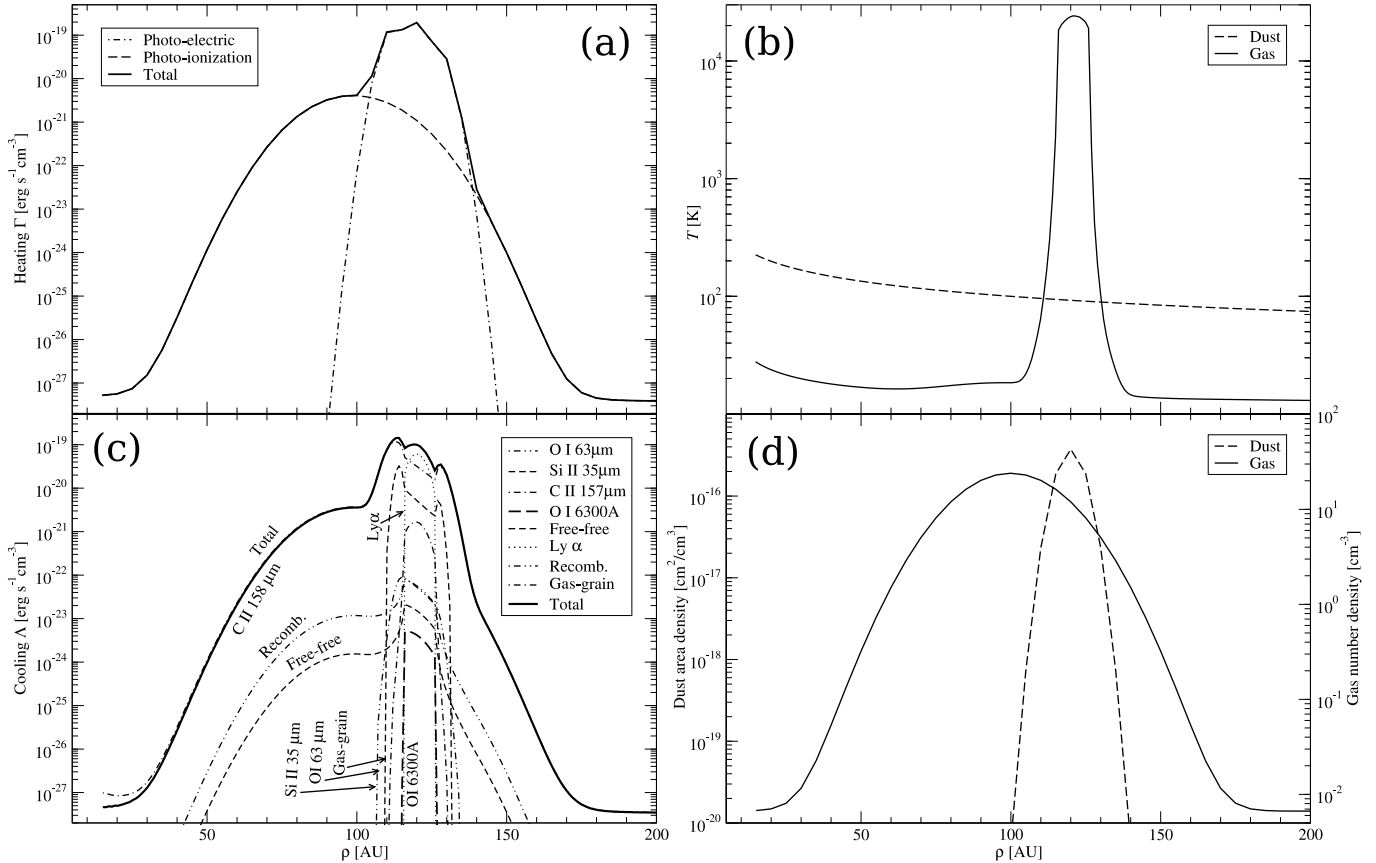}
\caption{Similar to Fig. \ref{fig:bPic_vs_rho}, but 
for the fiducial model. Due to the higher $n_{\mathrm{dust}}/n_{\mathrm{gas}}$
value at the dust peak, the gas temperature there experiences a spike
and rises beyond $5\,000$\,K (see \S\,\ref{sec:Fid_Temp} for discussion).}
\label{fig:vs_Rho}
\end{center}
\end{figure*}

Fig. \ref{fig:vs_Rho} shows the gas and dust temperature obtained in
our fiducial model, as well as various cooling and heating rates.
The gas is thermally decoupled from the dust, as the low gas-grain
collision rate is not able to equilibrate the two population.  We
  find that PE heating dominates the gas heating where dust density is
  high, and that PI heating dominates where dust density is relatively
  low.  Over most of the disk, cooling is dominated by the \ion{C}{2}
157.7 $\mu$m line. However, as gas temperatures rises above $\sim
100$\,K, the cooling flux of \ion{C}{2} saturates as the upper
occupation number $n_1$ approaches the LTE limiting value, $n_1
\rightarrow g_1/(g_0 + g_1) n_0$. Other cooling lines come into
importance, including the \ion{Si}{2} 34.8\,$\mu$m line, the
Ly\,$\alpha$ line, and gas-grain collisions.  The \ion{O}{1} lines,
estimated to be important by BW07, are insignificant in our model, due
to NLTE effects.

Fig.~\ref{fig:vs_Rho} shows the gas temperature at the peak of the
dust distribution to rise as high as $2.5\times10^4$\,K. Since
\textsc{ontario} does not include some cooling processes becoming important
above $5\,000$\,K (like \ion{C}{1} $\lambda$9849, \ion{C}{2} $\lambda$2324,
\ion{Fe}{2} 1.26\,$\mu$m, etc.\ \citealt{hol89}), this is an overestimate.\footnote{
However, our tests indicate that at the low gas density
assumed here, these cooling mechanisms are negligible even at
temperatures as high as $10^5$\,K.} There is a possibility, however,
that the temperature is high enough for the gas to thermally
evaporate. Let the criterion be that the sound speed (for Carbon) becomes
comparable to the orbital escape velocity:

\begin{equation}
T_{\mathrm{max}} = 1.7\times10^4 \,\left(\frac{r}{100\,\mathrm{AU}}\right)\,\mathrm{K}.
\end{equation}
As this temperature limit is below our (over-)estimate, we cannot rule
out gas evaporation from the disk; a more detailed study, outside the scope of
the present paper, would be required.

\subsubsection{Cooling line fluxes}
\label{sec:Fid_Cool}

Table~\ref{table:CoolFlux} lists cooling line luminosities integrated
over the whole disk for the fiducial case. \ion{C}{2} 157.7\,$\mu$m
remains the most luminous line, with \ion{Si}{2} 34.8$\mu$m following
within an order of magnitude, while \ion{O}{1} lines are very
underluminous compared to previous study \citep{bes07} due to NLTE
effects.  Most of the line flux is generated in region of highest gas
and dust densities: 93\,\% of the total
\ion{C}{2} 157.7\,$\mu$m flux arises in
the region extending radially from 100 to 140\,AU and vertically up to
20\,AU.

For \ion{C}{2}, which is close to LTE, the line flux is dominated by
collisional processes (eq.~[\ref{eqn:SECoolFlux}] $\approx$
eq.~[\ref{eqn:SECoolRate}]), which in turn is determined by the
heating processes (in this case, photoelectric heating). Compared to
the $\beta$\,Pic case, the fiducial case has a higher gas temperature
in the dust maximum and consequently a smaller photoelectric heating
rate. This explains largely the lower \ion{C}{2} flux in the latter.
This difference is further compounded by enhanced photoionization
heating in the $\beta$\,Pic case due to a larger amount of gas closer
to the star.  \ion{O}{1}, on the other hand, experiences an electron
density that falls far below its critical density for LTE and its line
flux arises largely from excitation by stellar photons (radiative
pumping) and not collisional excitation which leads to
cooling. This is reflected in the third column of
  Table~\ref{table:CoolFlux}, which shows that the cooling fraction
  (fraction of the line flux responsible for cooling the gas, the
  ratio of eq.~[\ref{eqn:SECoolRate}] to eq.~[\ref{eqn:SECoolFlux}])
  for \ion{O}{1} is $0.06 \ll 1$, while it is $0.96$ for \ion{C}{2}.

\begin{table}[ht]
\caption{Fiducial case cooling line luminosities}
\centering
\begin{tabular}{l l l c}
\hline
\hline
Line & Luminosity & Flux at 20\,pc & Cooling\\ & [$L_{\odot}$] &
[erg\,s$ ^{-1}$\,cm$^{-2}$] &Fraction\footnote{Fraction of line
luminosity from collisional processes (the ratio of eq.~[\ref{eqn:SECoolRate}] to eq.~[\ref{eqn:SECoolFlux}]); the remainder is due to
radiative pumping and does not contribute to cooling of the gas.}\\
\hline
\ion{C}{2} 157.7\,$\mu$m & $2.3\times 10^{-8}$  & $1.8\times 10^{-15}$ & 0.96\\
\ion{O}{1}  44.1\,$\mu$m & $3.5\times 10^{-16}$ & $2.8\times 10^{-23}$&$\rceil$\\
\ion{O}{1}  63.2\,$\mu$m & $1.5\times 10^{-10}$  & $1.2\times 10^{-17}$&0.06\footnote{Total \ion{O}{1} cooling fraction summed over all 3 transitions.}\\
\ion{O}{1} 145.5\,$\mu$m & $1.4\times 10^{-11}$ & $1.1\times 10^{-18}$&$\rfloor$\\
\ion{Si}{2} 34.8\,$\mu$m & $3.2\times 10^{-9}$  & $2.6\times 10^{-16}$&0.63\\
\hline
\end{tabular}
\tablecomments{Cooling line luminosities and fluxes measured  at 20\,pc, 
as predicted by \textsc{ontario} for the fiducial model.}
\label{table:CoolFlux}
\end{table}

\subsubsection{Ionization profile}
\label{sec:Fid_Ion}

\begin{figure}
\begin{center}
   \includegraphics[width=\columnwidth]{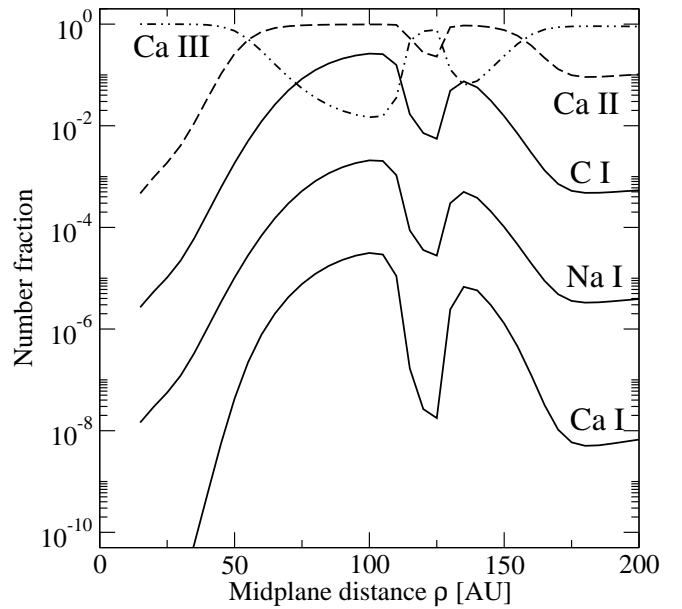} 
\caption{Midplane neutral (unbroken line) and ionization (dashed for one time ionized, and
dash-dotted for two times ionized) fractions of three selected
elements: Ca, Na, and C.}  \label{fig:radion}
\end{center}
\end{figure}

\begin{figure}
\begin{center}
   \includegraphics[width=\columnwidth]{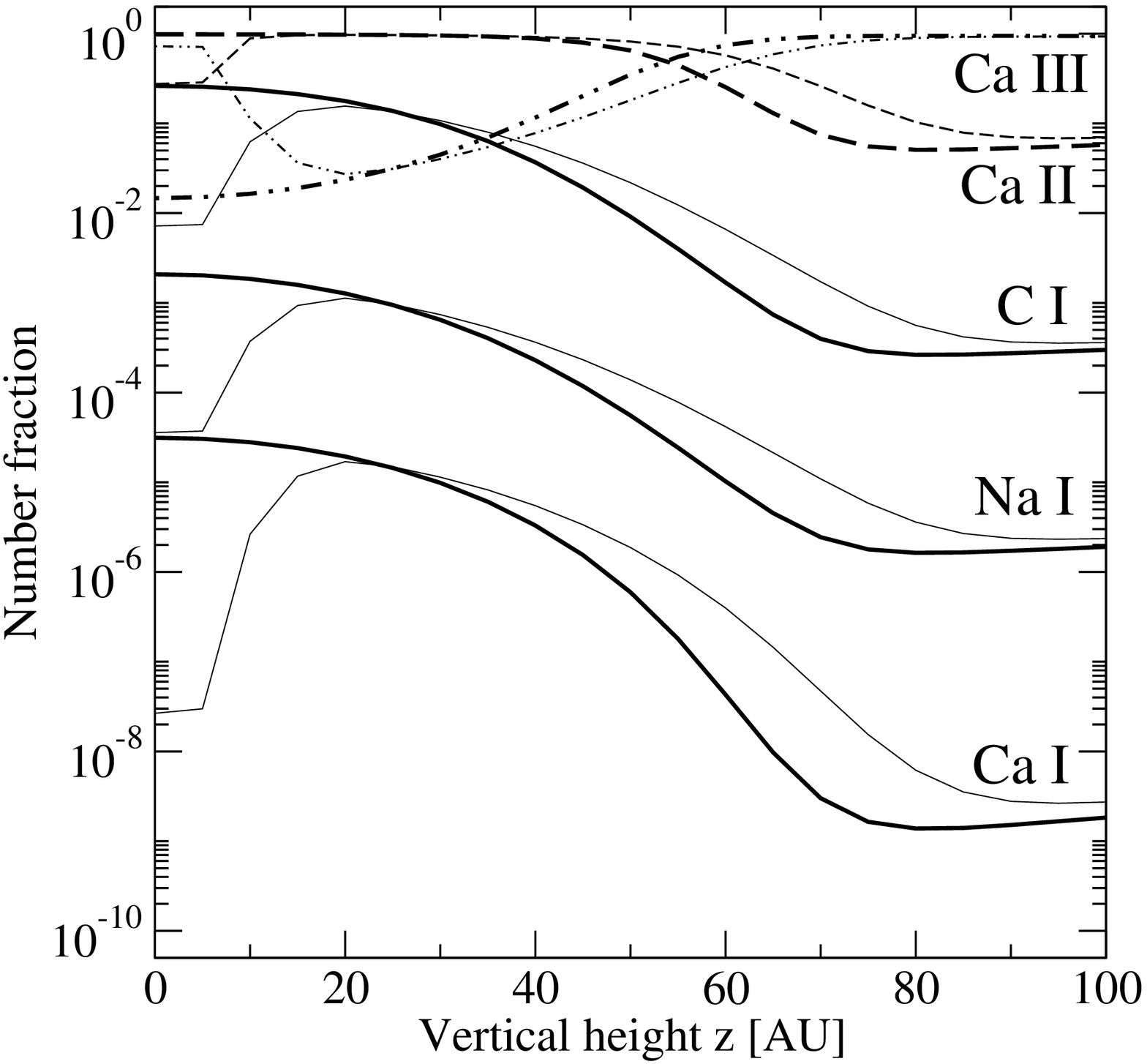} 
\caption{As in Fig.~\ref{fig:radion}, but for the vertical distribution at 100\,AU (thick line)
 and 120\,AU (thin line).}  \label{fig:vertion}
\end{center}
\end{figure}

In our fiducial model, the metallic gas is strongly ionized by the
central star.  Figs.~\ref{fig:radion}\,\&\,\ref{fig:vertion}\ show the
neutral fractions of Ca, Na, and C, the main contributers to the
electron density. We also show the fractions of \ion{Ca}{2} and
\ion{Ca}{3} since a significant fraction of Ca is ionized a second 
time.

The ionization profile is explained by a competition between stellar
photoionization and recombination. Rate for the former falls off as
one moves away from the star, while rate for the latter scales with
electron density (and weakly depends on temperature through the recombination
coefficient). So the neutral fraction is the lowest near the star and
it rises upward until it reaches a maximum around 100\,AU. Outward of
this distance, the sharp drop of gas density reduces the recombination
rate faster than the decrease of the ionization rate, resulting again
in a more ionized gas.
The neutral fractions exhibit a narrow dip at around 120\,AU. This is
due to the temperature peak (as shown in Fig.~\ref{fig:vs_Rho}), which
in turn is due to the dust peak at that location.
As is seen in Fig.~\ref{fig:vertion}, the neutral fractions decrease
monotonically away from the midplane due to the reduction in
recombination in lower densities.

\subsubsection{Optical/UV absorption and emission lines}
\label{sec:Fid_Spectr}

Observationally, an advantage of the IR cooling lines is that the
stellar photosphere is relatively dark at those wavelengths, making
faint emission lines more easily detectable. Should the disk be 
spatially resolvable or be observed edge on, strong transitions
in the optical/UV might still be easier to detect. Indeed, gas around
$\beta$\,Pic was first detected in absorption \citep{sle75,hob85},
and then in spatially resolved emission from light scattered in 
atomic resonance lines \citep[\ion{Na}{1} D$_{2,1}$; ][]{olo01}; apart
from the tentative
detection of \ion{C}{2} 157.7\,$\mu$m emission \citep{kam03}, cooling
lines have yet to be observed from the disk. With this in mind, we have
computed equivalent widths for absorption lines assuming an edge-on disk
(Table~\ref{tbl:Absorp}), and total luminosities for light scattered in lines with
strong transitions (Table~\ref{tbl:Emissions}). Resolved observations of the 
\ion{Ca}{2} K absorption line around $\beta$\,Pic \citep{cra98} show it
to be $\sim$ 2\,km\,s$^{-1}$ wide. We assume absorption from the `stable
component' of other species to be similarly broadened, resulting in 
increasing saturation for lines with equivalent widths approaching
2\,km\,s$^{-1}$.

\begin{table}[ht]
\caption{Fiducial Case Dominant Absorption Lines}
\centering
\begin{tabular}{l c c}
\hline
\hline
Species& $\lambda_{\mathrm{vacuum}}$\footnote{Only  lines with $\lambda \geq 3300$ \AA\, are presented.}&Eq.\ Width\footnote{OT - optically thick line, defined to have an equivalent width
that corresponds to a velocity $\gtrsim 2$\,km\,s$^{-1}$.}\\
&[\AA]&[m\AA]\\
\hline
\ion{Na}{1}&5891.58&$0.86$\\
\ion{Na}{1}&5897.56&$0.43$\\
\ion{Ca}{2}&3934.78&OT\\
\ion{Ca}{2}&3969.59&OT\\
\ion{Ti}{2}&3350.37&$0.82$\\
\ion{Ti}{2}&3362.18&$0.59$\\
\ion{Ti}{2}&3373.77&$0.51$\\
\ion{Ti}{2}&3384.74&$0.39$\\
\ion{Mn}{2}&3442.97&$0.15$\\
\ion{Fe}{1}&3441.59&$0.31$\\
\ion{Fe}{1}&3720.99&$0.88$\\
\ion{Fe}{1}&3735.93&$0.36$\\
\ion{Fe}{1}&3861.01&$0.53$\\
\hline
\end{tabular}
\label{tbl:Absorp}
\end{table}

\begin{table}[ht]
\caption{Fiducial Case Dominant Emission Lines}
\centering
\begin{tabular}{l c c c}
\hline
\hline
Species & $\lambda_{\mathrm{vacuum}}$\footnote{Only  lines with $\lambda \geq 3300$\,\AA\ are presented.}& Luminosity & Flux at 20\,pc \\
        && $L_{\odot}$ & [erg\,s$ ^{-1}$\,cm$^{-2}$]\\
\hline

\ion{C}{1} &9826.8\,\AA&$1.5\times 10^{-8}$&$1.2\times 10^{-15}$\\
\ion{C}{1} &9853.0\,\AA&$4.4\times 10^{-8}$&$3.5\times 10^{-15}$\\

\ion{C}{2} &157.7\,$\mu$m&$2.3\times 10^{-8}$&$1.8\times 10^{-15}$\\

\ion{O}{1} &63.2\,$\mu$m&$1.5\times 10^{-10}$&$1.2\times 10^{-17}$\\

\ion{Na}{1} &5897.6\,\AA&$7.6\times 10^{-8}$&$6.1\times 10^{-15}$\\
\ion{Na}{1} &5891.6\,\AA&$1.5\times 10^{-7}$&$1.2\times 10^{-14}$\\

\ion{Al}{1} &3945.1\,\AA&$2.2\times 10^{-10}$&$1.8\times 10^{-17}$\\
\ion{Al}{1} &3962.6\,\AA&$4.4\times 10^{-10}$&$3.5\times 10^{-17}$\\

\ion{Si}{2} &34.8\,$\mu$m&$3.2\times 10^{-9}$&$2.6\times 10^{-16}$\\

\ion{S}{1} &25.2\,$\mu$m&$2.4\times 10^{-10}$&$1.9\times 10^{-17}$\\

\ion{Ca}{2} * &3934.8\,\AA&$9.4\times 10^{-6}$&$7.6\times 10^{-13}$\\
\ion{Ca}{2} * &3969.6\,\AA&$7.0\times 10^{-6}$&$5.6\times 10^{-13}$\\
\ion{Ca}{2} &7293.5\,\AA&$3.4\times 10^{-7}$&$2.8\times 10^{-14}$\\
\ion{Ca}{2} &7325.9\,\AA&$3.2\times 10^{-7}$&$2.5\times 10^{-14}$\\

\ion{Ti}{2} &4534.5\,\AA&$4.2\times 10^{-10}$&$3.4\times 10^{-17}$\\
\ion{Ti}{2} &4983.1\,\AA&$4.6\times 10^{-10}$&$3.7\times 10^{-17}$\\

\ion{Ti}{2} &3350.4\,\AA&$1.6\times 10^{-7}$&$1.3\times 10^{-14}$\\
\ion{Ti}{2} &3362.2\,\AA&$1.1\times 10^{-7}$&$8.8\times 10^{-15}$\\
\ion{Ti}{2} &3373.8\,\AA&$1.1\times 10^{-7}$&$8.8\times 10^{-15}$\\
\ion{Ti}{2} &3384.7\,\AA&$8.4\times 10^{-8}$&$6.8\times 10^{-15}$\\

\ion{Cr}{2} &8002.3\,\AA&$7.4\times 10^{-8}$&$5.9\times 10^{-15}$\\
\ion{Cr}{2} &8127.5\,\AA&$5.8\times 10^{-8}$&$4.7\times 10^{-15}$\\
\ion{Cr}{2} &8231.9\,\AA&$4.7\times 10^{-8}$&$3.8\times 10^{-15}$\\

\ion{Mn}{1} &4034.2\,\AA&$2.2\times 10^{-9}$&$1.8\times 10^{-16}$\\
\ion{Mn}{1} &4031.9\,\AA&$3.0\times 10^{-9}$&$2.4\times 10^{-16}$\\

\ion{Mn}{2} &3443.0\,\AA&$3.5\times 10^{-8}$&$2.8\times 10^{-15}$\\
\ion{Mn}{2} &3461.3\,\AA&$1.8\times 10^{-8}$&$1.5\times 10^{-15}$\\

\ion{Fe}{1} &3721.0\,\AA&$1.4\times 10^{-7}$&$1.1\times 10^{-14}$\\
\ion{Fe}{1} &3735.9\,\AA&$8.6\times 10^{-8}$&$6.9\times 10^{-14}$\\
\ion{Fe}{1} &3821.5\,\AA&$9.9\times 10^{-8}$&$7.9\times 10^{-15}$\\
\ion{Fe}{1} &3861.0\,\AA&$1.3\times 10^{-7}$&$1.1\times 10^{-14}$\\

\ion{Fe}{2} &12570.9\,\AA&$1.6\times 10^{-7}$&$1.3\times 10^{-14}$\\
\ion{Fe}{2} &13209.9\,\AA&$4.3\times 10^{-8}$&$3.5\times 10^{-15}$\\
\ion{Fe}{2} &16440.0\,\AA&$4.1\times 10^{-8}$&$3.3\times 10^{-15}$\\
\ion{Fe}{2} &25.99\,$\mu$m&$6.1\times 10^{-8}$&$4.9\times 10^{-15}$\\

\ion{Ni}{2} &6668.6\,\AA&$1.9\times 10^{-8}$&$1.5\times 10^{-15}$\\
\ion{Ni}{2} &7379.9\,\AA&$8.0\times 10^{-8}$&$6.4\times 10^{-15}$\\
\ion{Ni}{2} &7413.7\,\AA&$3.1\times 10^{-8}$&$2.5\times 10^{-15}$\\

\hline
\end{tabular}
\tablecomments{The 
two \ion{Ca}{2} lines indicated by (*) are optically thick. While our
code includes the scattering of stellar flux from the star to the
computational bin, it does not account for the scattering of gas
emissions along the line of sight of the observer. Therefore, the
computed values may significantly overestimate the luminosities in the
2 saturated calcium lines.}
\label{tbl:Emissions}
\end{table}

\subsection{Parametric survey}
\label{sec:Discussion}

To detect circumstellar gas in absorption requires a special geometry,
and to detect light scattered in gas most likely requires a resolved
disk. A comprehensive survey for tenuous gas in debris disks is
therefore best served by the IR cooling lines.  In this section, we
study how the cooling line luminosities depend on various disk
parameters, focusing on the \ion{C}{2} 157.7\,$\mu\mathrm{m}$ and
\ion{O}{1} 63.2\,$\mu\mathrm{m}$ lines, since these fall in the
wavelength window of the Herschel telescope.  The parameter study is
obtained by modifying one parameter at a time while keeping all other
parameters at their fiducial values. Disk parameters and their
  ranges (provided for easy reference in Table~\ref{table:Parameters})
  that we cover include:

\begin{enumerate}

\item peak gas density. Range covered is from 0.01 $\beta$\,Pic
  density ($2.4\times 10^{-1}$\,cm$^{-3}$) to $\sim
  2.7\times10^3$\,cm$^{-3}$, the latter corresponds to a column
  density above which the \ion{C}{2} 157.7\,$\mu$m line becomes
  radially optically thick.  For the spatial distribution we adopt,
  these densities correspond (through
    eq.~[\ref{eqn:n_GasDust_Fid}] with fiducial parameters) to a
  total gas masses of $2.1\times10^{-6} M_\oplus$ and
  $3.3\times10^{-2} M_{\oplus}$, respectively.

\item total dust mass. We sample dust mass from 1
zodiac (taken as $M_{\mathrm{dust}} = 10^{-3}$\,$M_{\oplus}$; with
fiducial disk parameters, this corresponds to $L_{\mathrm{dust}}/L_*
\sim 10^{-6}$) to 3\,$M_{\oplus}$, at which point the dust
becomes radially optically thick (with $L_{\mathrm{dust}}/L_* \sim 3
\times 10^{-3}$).

\item spatial distributions of the gas and dust components, including 
location of peak density, radial FWHM and vertical scale height,
spanning about an order of magnitude in each parameter.

\item elemental abundances. We vary carbon and oxygen abundances, respectively, 
from solar to 20$\times$ solar, to account for the type of abundance
anomaly detected in the $\beta$\,Pic disk.

\item stellar spectral type. We study main sequence stars with 
effective photosphere temperatures ranging from 6\,500\,K to
10\,000\,K. For cooler stars, molecular chemistry is likely to be
important, rendering the \textsc{ontario} code invalid. 
We interpolate between the ZAMS models in Table 15.14
in Allen's Astrophysical Quantities \citep{cox00} to obtain the
stellar luminosities and sizes.
\end{enumerate}

\begin{table}[ht]
\caption{Parametric survey: overview of sampled parameters}
\centering
\begin{tabular}{l c c c}
\hline
\hline
Parameter sampled & Min & Max & \# of bins \\
\hline
Gas peak density & $2.4\times 10^{-1}$\,cm$^{-3}$ & $2.7\times 10^{3}$\,cm$^{-3}$ & 9 \\
(total gas mass) & ($2.1\times 10^{-6}$\,$M_\oplus$) & ($3.3\times 10^{-2}$\,$M_\oplus$) & \\
Dust mass & 10$^{-3}$\,$M_\oplus$ & 3\,$M_\oplus$ & 8 \\
Gas peak location & 30\,AU & 190\,AU & 33 \\
Gas FWHM & 5\,AU & 95\,AU & 19 \\
Gas scale height & 0.05 & 0.95 & 19 \\
Dust peak location & 30\,AU & 190\,AU & 33 \\
Dust FWHM & 5\,AU & 95\,AU & 19 \\
Dust scale height & 0.05 & 0.95 & 19 \\
Spectral types (T$_\mathrm{eff}$) &  6\,500\,K & 10\,000\,K & 8 \\
Elemental abundances & solar & $20\times$ solar & \\
\hline
\end{tabular}
\label{table:Parameters}  
\end{table}

\begin{figure}
\begin{center}
      \includegraphics[width=\columnwidth]{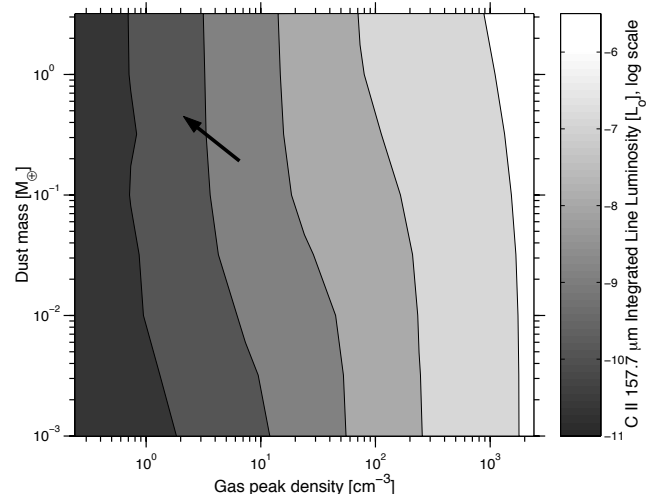} 
\caption{\ion{C}{2}
      157.7\,$\mu$m integrated line luminosity as a function of peak
      gas density and total dust mass. Contours are separated by a
      factor of 10. The arrow shows the direction of increasing temperature.
}
      \label{fig:CoolC_vs_DustGas}
\end{center}
\end{figure}

\subsubsection{Gas and dust densities}

\begin{figure}
\begin{center}
      \includegraphics[width=\columnwidth]{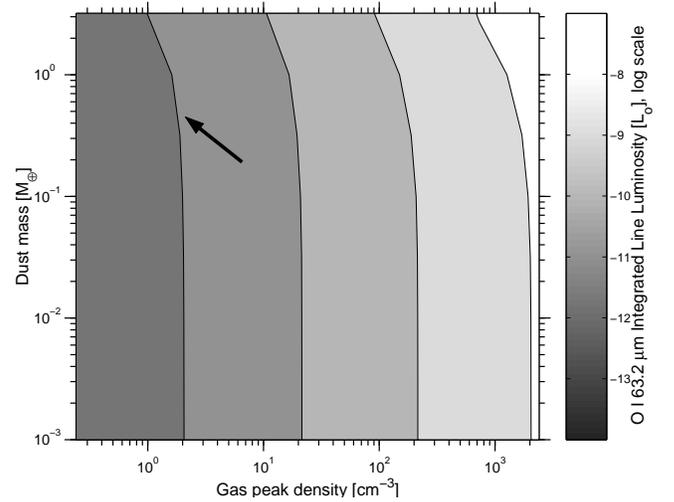} 
\caption{\ion{O}{1}
      63.2\,$\mu$m integrated line luminosity as a function of peak
      gas density and total dust mass. A 10-level \ion{O}{1} atom is
      used in the computation \edII{(see Appendix for details)}. The line luminosity is largely independent of
	dust mass and scales roughly linearly with gas density.} \label{fig:CoolO_vs_DustGas}
\end{center}
\end{figure}

We show the integrated luminosities from the \ion{C}{2} 157.7\,$\mu$m
and the \ion{O}{1} 63.2\,$\mu$m lines as functions of gas peak density
and dust mass in Figs.~\ref{fig:CoolC_vs_DustGas} \&
\ref{fig:CoolO_vs_DustGas}, respectively. 
In systems with little dust, gas is primarily heated through
photoionization, so the \ion{C}{2} luminosity scales almost linearly
with the amount of gas.
As more dust is added, photoelectric heating takes over and the
\ion{C}{2} flux starts to increase with both dust and gas density
(eq.~[\ref{eq:PEguess}]).  For very massive dust disks, gas
temperature is so high that the \ion{C}{2} flux becomes saturated and
once again loses its dependence on the dust mass. 
Despite minor features, one can summarize the results in
Fig.~\ref{fig:CoolC_vs_DustGas} as that the \ion{C}{2} flux rises
roughly linearly with gas density.

\begin{figure}
\begin{center}
      \includegraphics[width=\columnwidth]{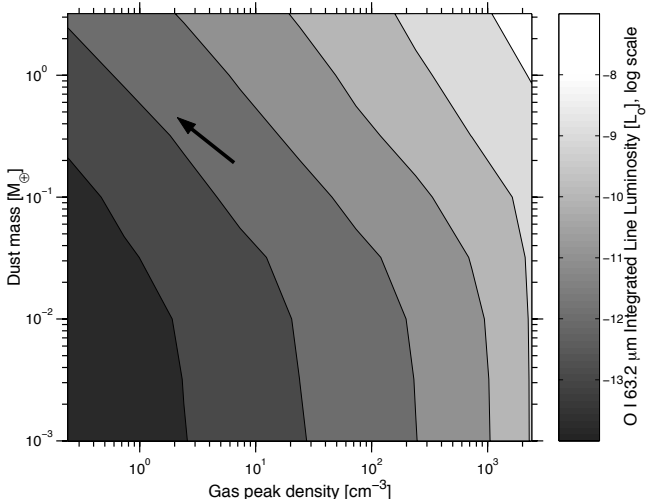} 
\caption{Similar to Fig. \ref{fig:CoolO_vs_DustGas} except that here 
	a 3-level \ion{O}{1} atom is adopted. This removes the possibility of
	radiative pumping. Line luminosities are much weaker and they depend
	on dust mass.
}  \label{fig:CoolO3_vs_DustGas}
\end{center}
\end{figure}

\begin{figure}
\begin{center}
      \includegraphics[width=\columnwidth]{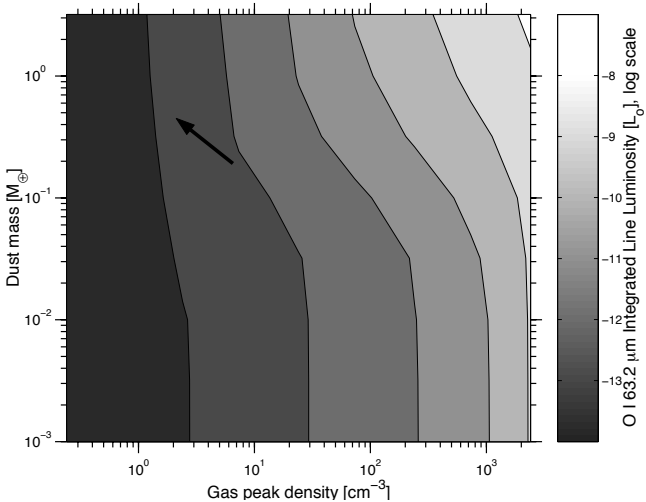} 
\caption{Similar to Fig. \ref{fig:CoolO_vs_DustGas} except that here 
	a 3-level \ion{O}{1} atom is adopted, and dust IR field is disabled.
}  \label{fig:CoolO3NoDustIR_vs_DustGas}
\end{center}
\end{figure}
Except for the most massive dust disks where SE begins to be
  dominated by the strong dust IR field, the contours for the
  \ion{O}{1} 63.2\,$\mu$m line flux appear independent of the dust
  mass for most of our disks (Fig.~\ref{fig:CoolO_vs_DustGas}).  This
  is explained by the fact that the line flux is almost completely
  dominated by radiative pumping, which masks any features from the
  collisional processes.  To remove this masking we repeat the
  computation but with the \ion{O}{1} atom having only the three
  lowest energy states (as opposed to the 10 levels in our normal
  calculation, \edII{discussed in greater detail in the Appendix}). 
  Data both with and without the dust IR field are
  presented in Figs.\ \ref{fig:CoolO3_vs_DustGas} and
  \ref{fig:CoolO3NoDustIR_vs_DustGas}, respectively.
  \edII{For dust disk masses $\lesssim$0.2--0.3\,M$_{\oplus}$,
the \ion{O}{1} line flux is completely dominated by radiative
pumping from the star, and neglecting this process can underestimate
the cooling line flux by more than a factor of 100.}  With radiative pumping disabled,
the results show similar dust mass dependence as that of the \ion{C}{2}
line. For more massive disks the flux from the cooling lines is dominated
by the population equilibrium determined by the dust IR field.
\begin{figure*}
\begin{center}
      \includegraphics[scale=0.28]{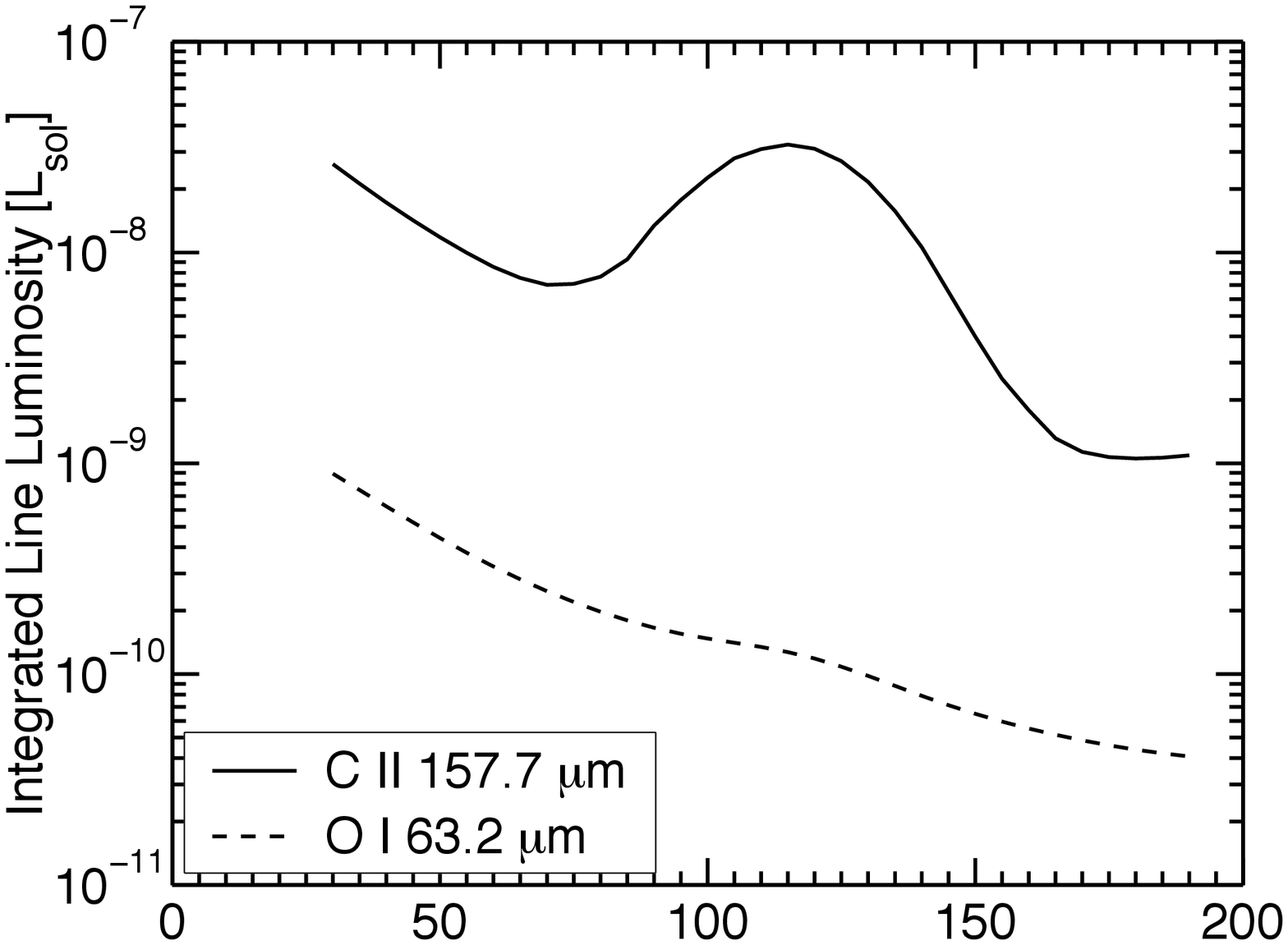}
      \includegraphics[scale=0.28]{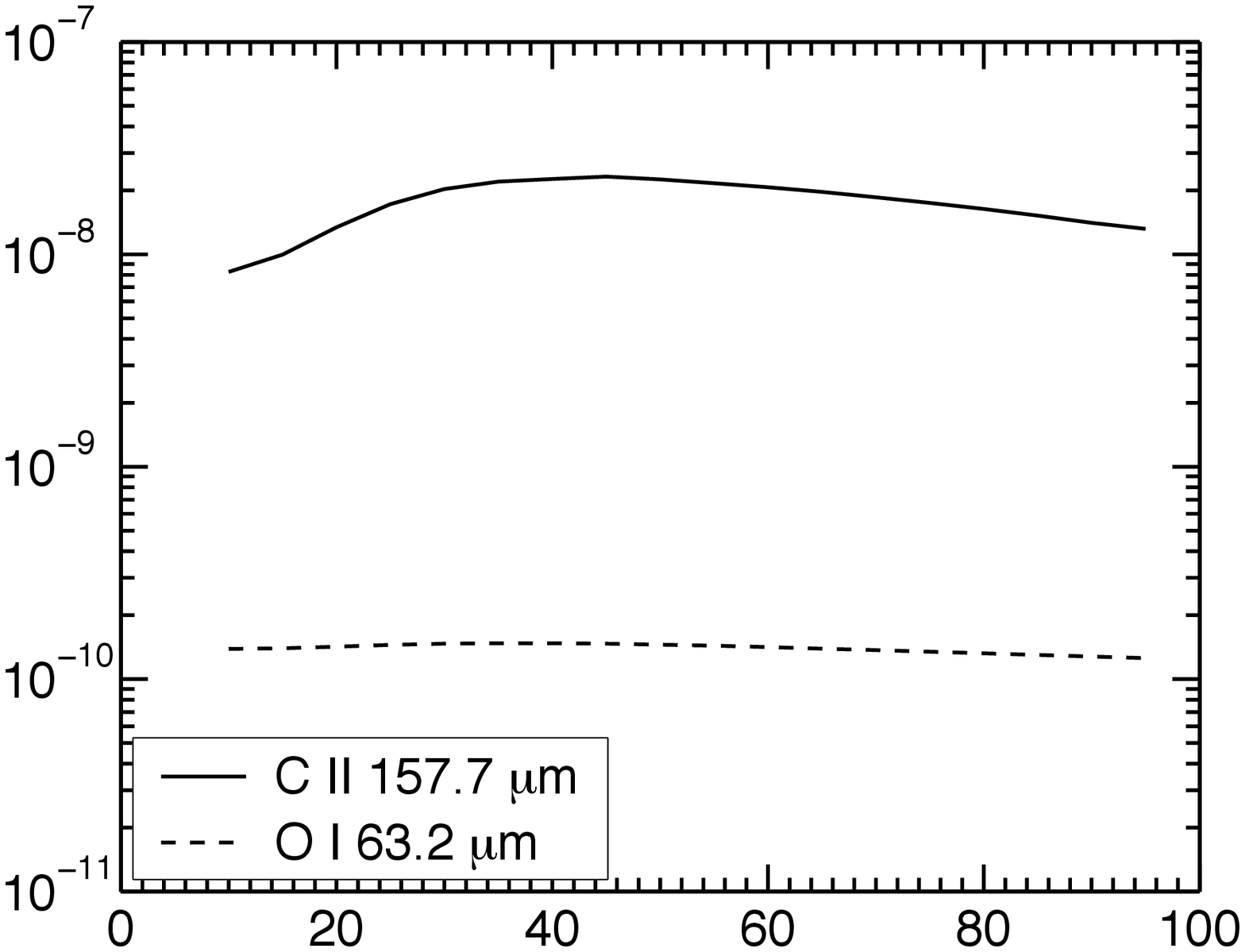}
      \includegraphics[scale=0.28]{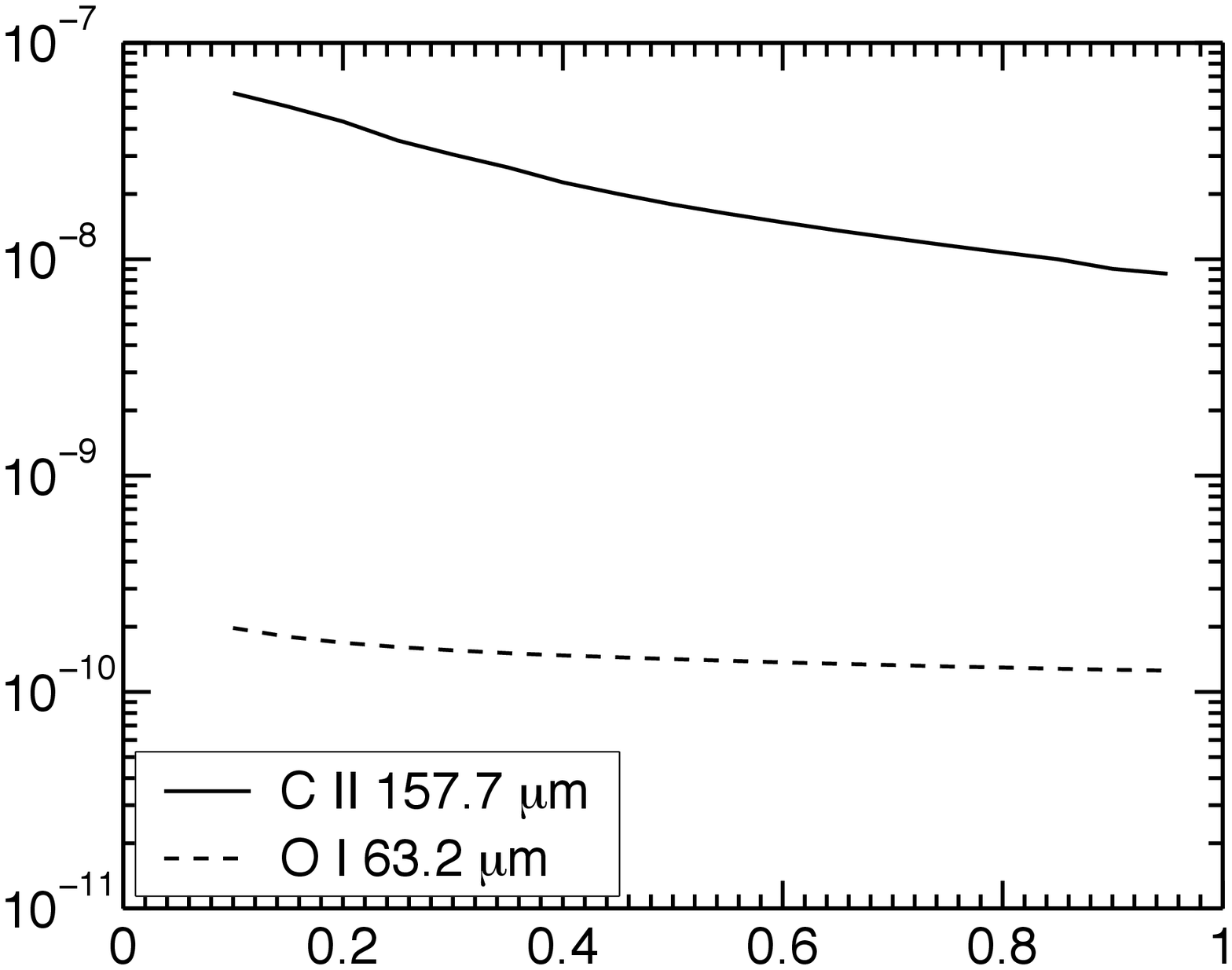} \\
      \includegraphics[scale=0.28]{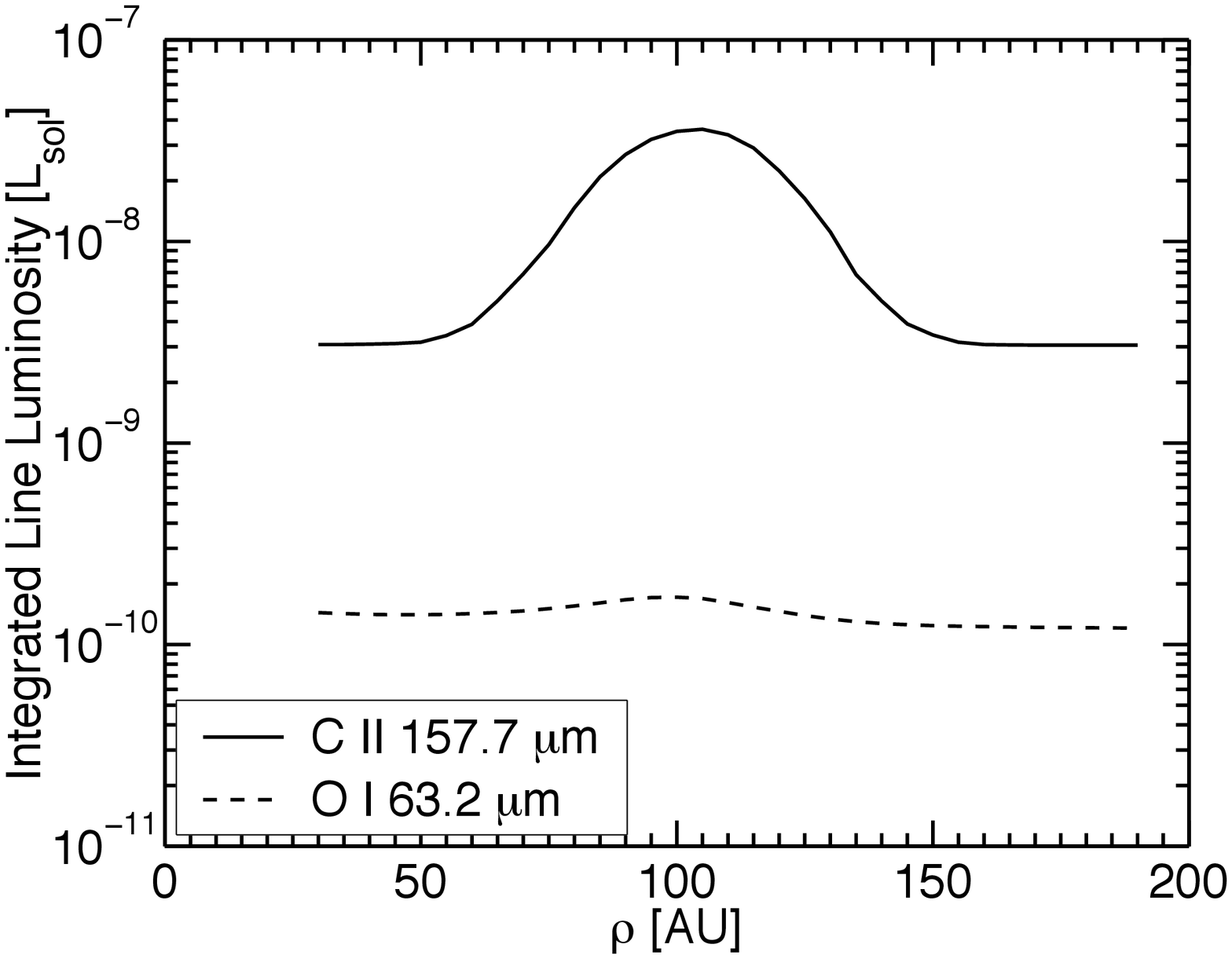}
      \includegraphics[scale=0.28]{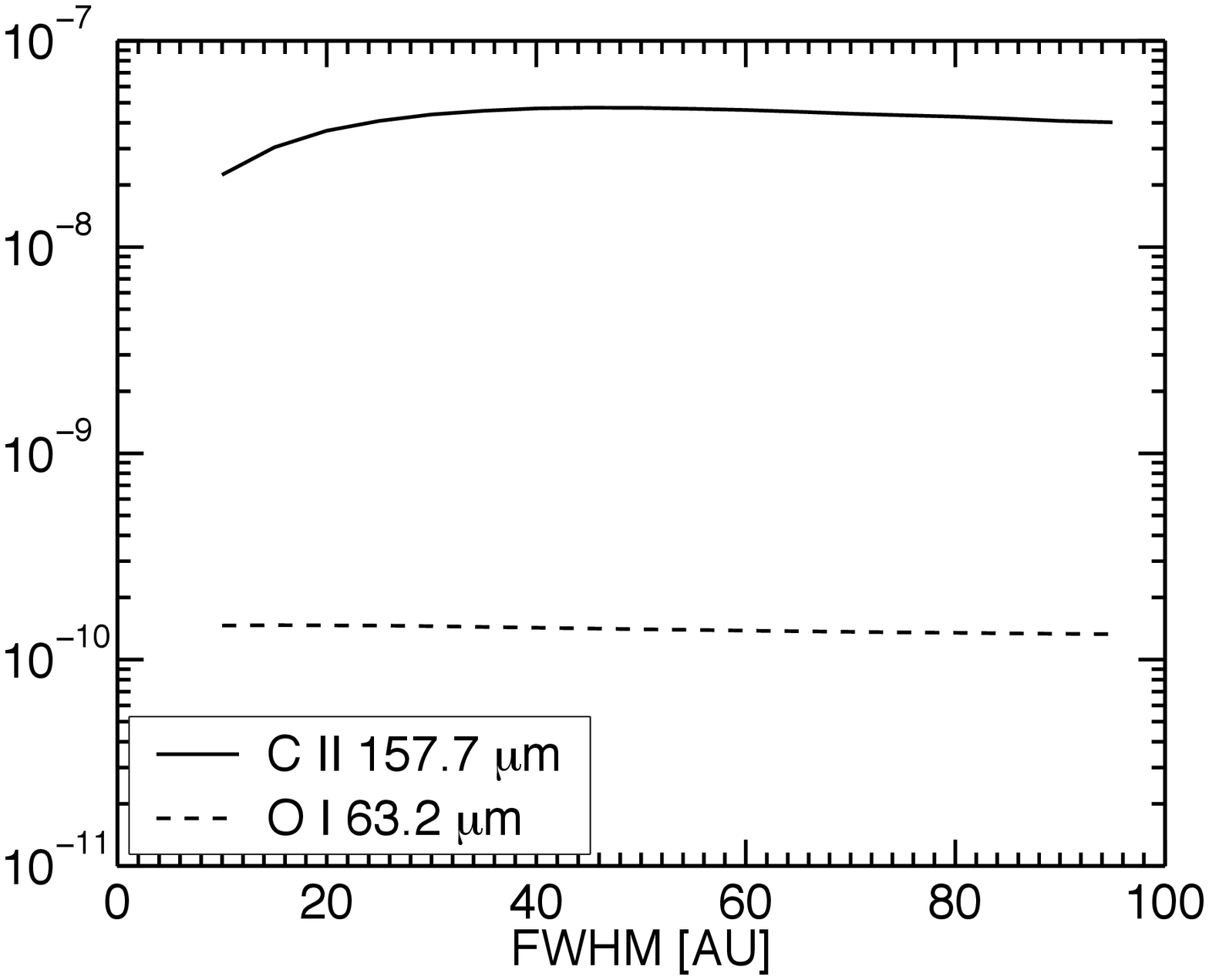}
      \includegraphics[scale=0.28]{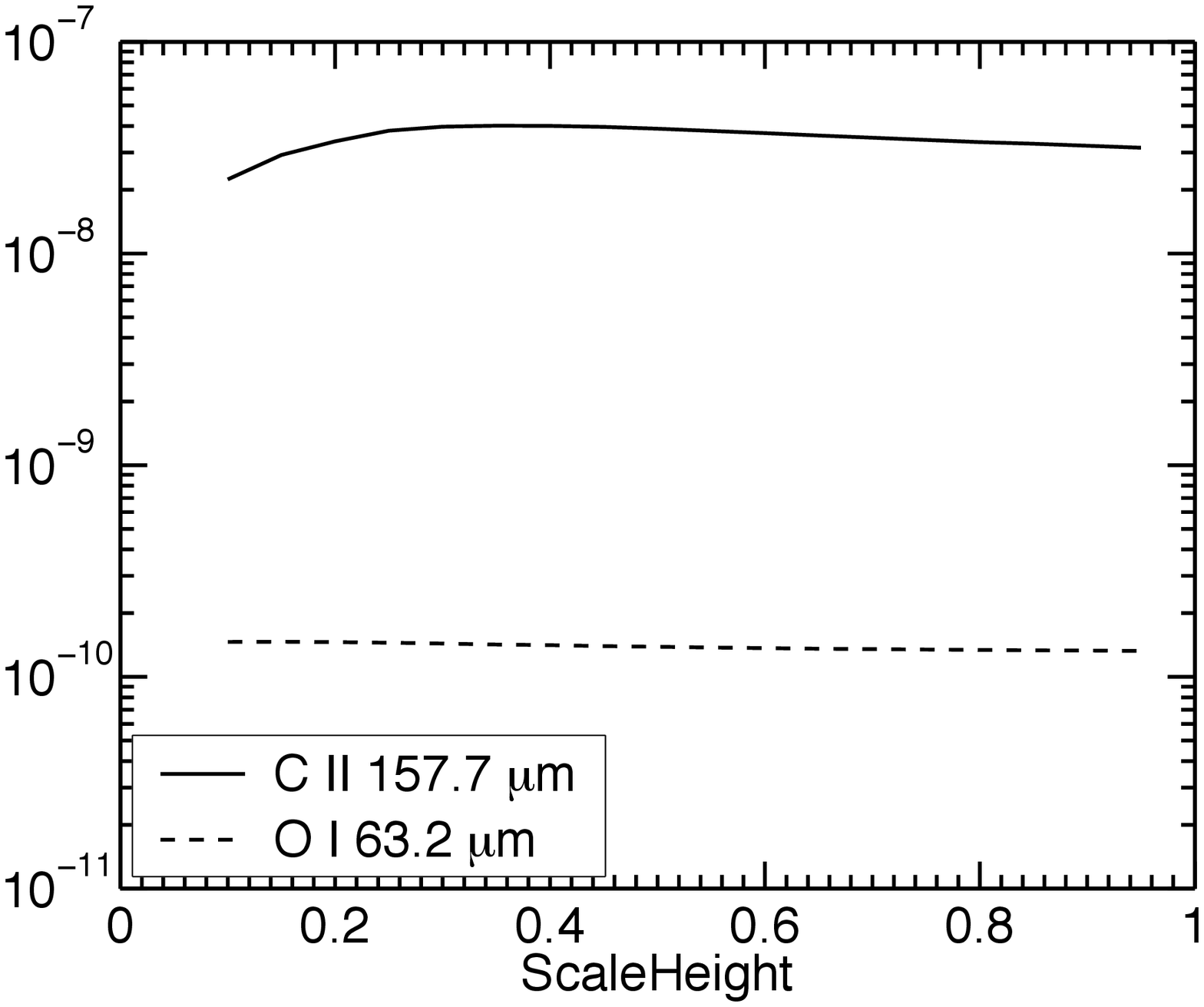}
      \caption{\ion{C}{2} 157.7\,$\mu$m
      and \ion{O}{1} 63.2\,$\mu$m integrated line luminosities as functions of
      gas (top) and dust (bottom) peak cylindrical radii (left panels), 
	their radial FWHM (middle panels) and their scale heights (right panels). 
One parameter is varied at a time. Only the left panels show strong dependence -- the 
line radiation is enhanced whenever the peak radii for the dust and the gas coincide, or when
bulk of the gas lies very close to the star.}
      \label{fig:Cool_vs_GasDustParams}
\end{center}
\end{figure*}

In Fig.~\ref{fig:Cool_vs_GasDustParams}, cooling luminosities are
presented as a function of a number of parameters that define the
geometry of the gas and dust distributions. 

Overall, the \ion{C}{2} 157.7\,$\mu$m luminosity is at the greatest
when peaks of the gas and dust distributions coincide, allowing for
maximum photoelectric heating, or when bulk of the gas is located
close to the star, allowing for maximum radiative pumping and
photoionization heating.  The \ion{O}{1} luminosity steadily increases
as the gas distribution is moved closer to the star as this line is
dominated by radiative pumping.  In contrast, the radial width and the
vertical scale height for the distributions do not affect line
luminosities appreciably. Besides having a minimal upregulating
  effect on the \ion{O}{1} 63.2\,$\mu$m line flux (as observed for the
  $\beta$\,Pic configuration), changes in the dust IR field due to
  varying dust profiles has no effect on the cooling line fluxes.

\subsubsection{Elemental abundances}

\begin{figure*}
\begin{center}
      \includegraphics[width=\columnwidth]{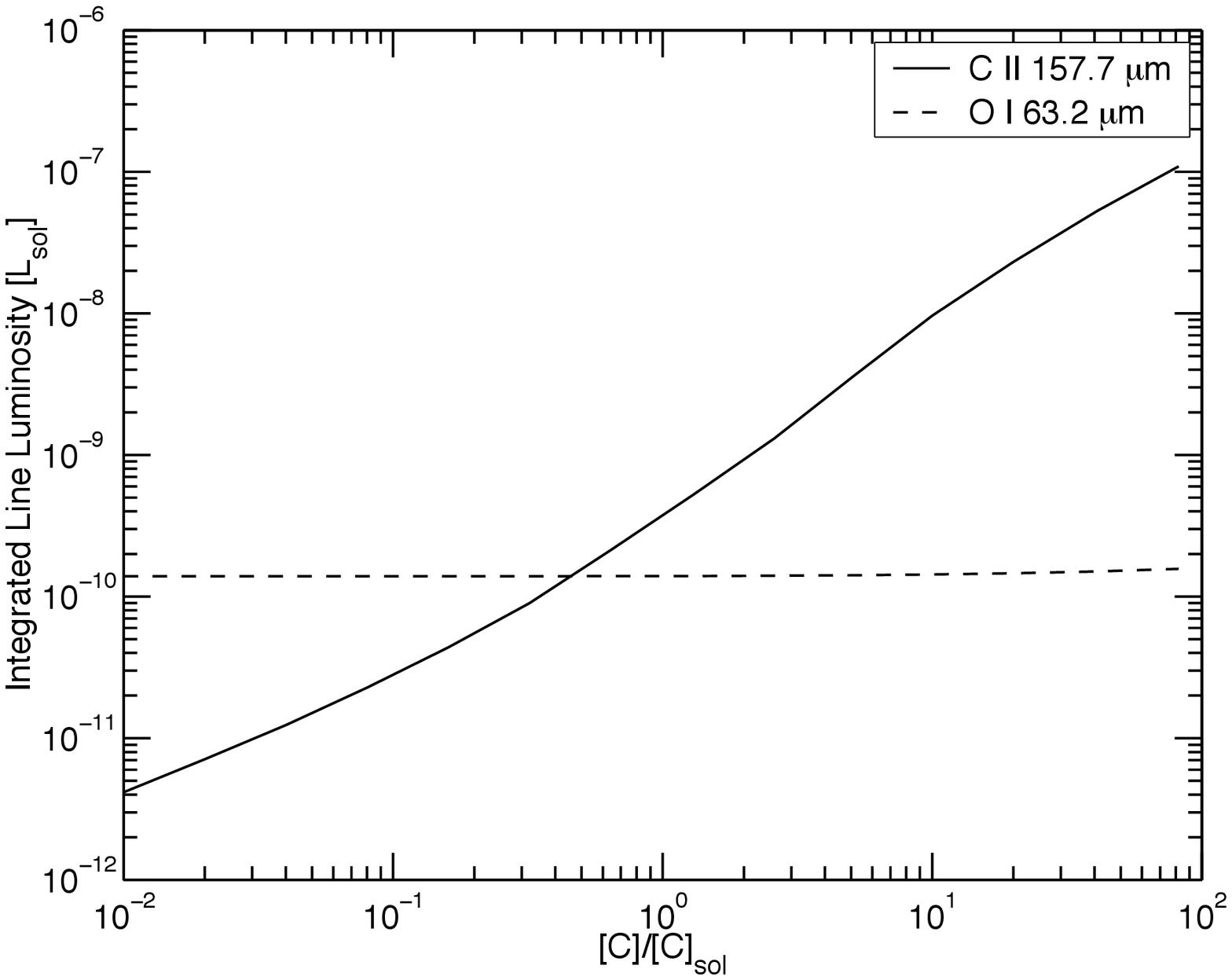}
      \includegraphics[width=\columnwidth]{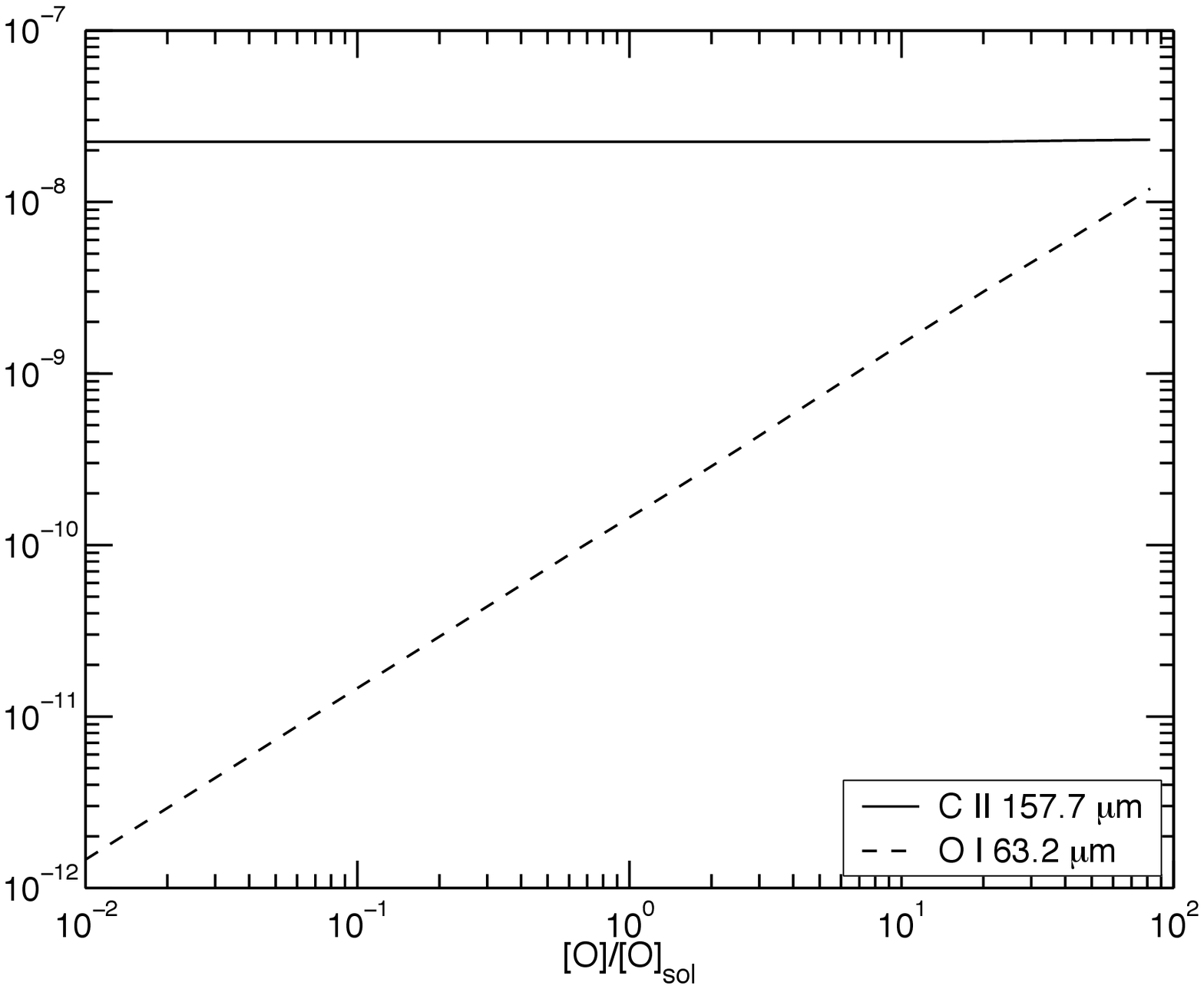}
      \caption{\ion{C}{2} 157.7\,$\mu$m and \ion{O}{1} 63.2\,$\mu$m integrated line
luminosities as function of C (left) and O (right) abundances. The former rises with carbon abundance, while
the latter with oxygen abundance, but the reasons behind are somewhat different (see text).}
   \label{fig:Cool_vs_Abund}
\end{center}
\end{figure*}

Since carbon is the dominant electron donor in the fiducial disk,
  increasing the carbon abundance increases $n_\mathrm{e}$. This, in
  turn, enhances the photoelectric heating as is demonstrated by
  eq.~[\ref{eqn:Je}].  The effects on the line fluxes are seen in
Fig.~\ref{fig:Cool_vs_Abund} (and Table~\ref{table:ModCoolFlux}). In
contrast, increasing the O abundance has no significant effect on the
thermal balance, besides from raising the \ion{O}{1} 63.2\,$\mu$m flux
by providing more atoms to be radiatively pumped.

We assume the gas to be poor in hydrogen. Raising the hydrogen
abundance to its solar value increases the gas-grain collisions which
cools the gas. This lowers the equilibrium gas temperature and 
slightly increases the photoelectric heating (and the corresponding 
\ion{C}{2} cooling flux). But the effect is not significant.
In fact, one can remove all elements from the gas except for C, O, Si
without causing much change in the cooling line luminosities 
(Table \ref{table:ModCoolFlux}). 

\begin{table}[ht]
\caption{Line luminosities as functions of elemental abundances}
\centering
\begin{tabular}{l c c}
\hline
\hline
Modification\footnote{Fiducial case inputs are used except for the parameter
mentioned.} &\ion{C}{2} 157.7\,$\mu$m&\ion{O}{1} 63.2\,$\mu$m\\
&[$L_{\odot}$] &[$L_{\odot}$]\\
\hline
Fiducial&$2.3\times 10^{-8}$&$1.3\times 10^{-10}$\\
$n_{\mathrm{C}}$ = $n_{\mathrm{C},\odot}$&$3.8\times 10^{-10}$&$1.2\times 10^{-10}$\\
$n_{\mathrm{O}}$ = 20 $n_{\mathrm{O},\odot}$&$2.3\times 10^{-8}$&$2.5\times 10^{-9}$\\
$n_{\mathrm{H}}$ = $n_{\mathrm{H},\odot}$&$2.6\times 10^{-8}$&$1.3\times 10^{-10}$\\
C, O, Si only\footnote{All elements except for C, O, Si and H are removed.}
&$2.3\times 10^{-8}$&$1.3\times 10^{-10}$\\
\hline
\label{table:ModCoolFlux}
\end{tabular}
\end{table}

\subsubsection{Stellar spectral type}

\begin{figure}
\begin{center}
      \includegraphics[width=\columnwidth]{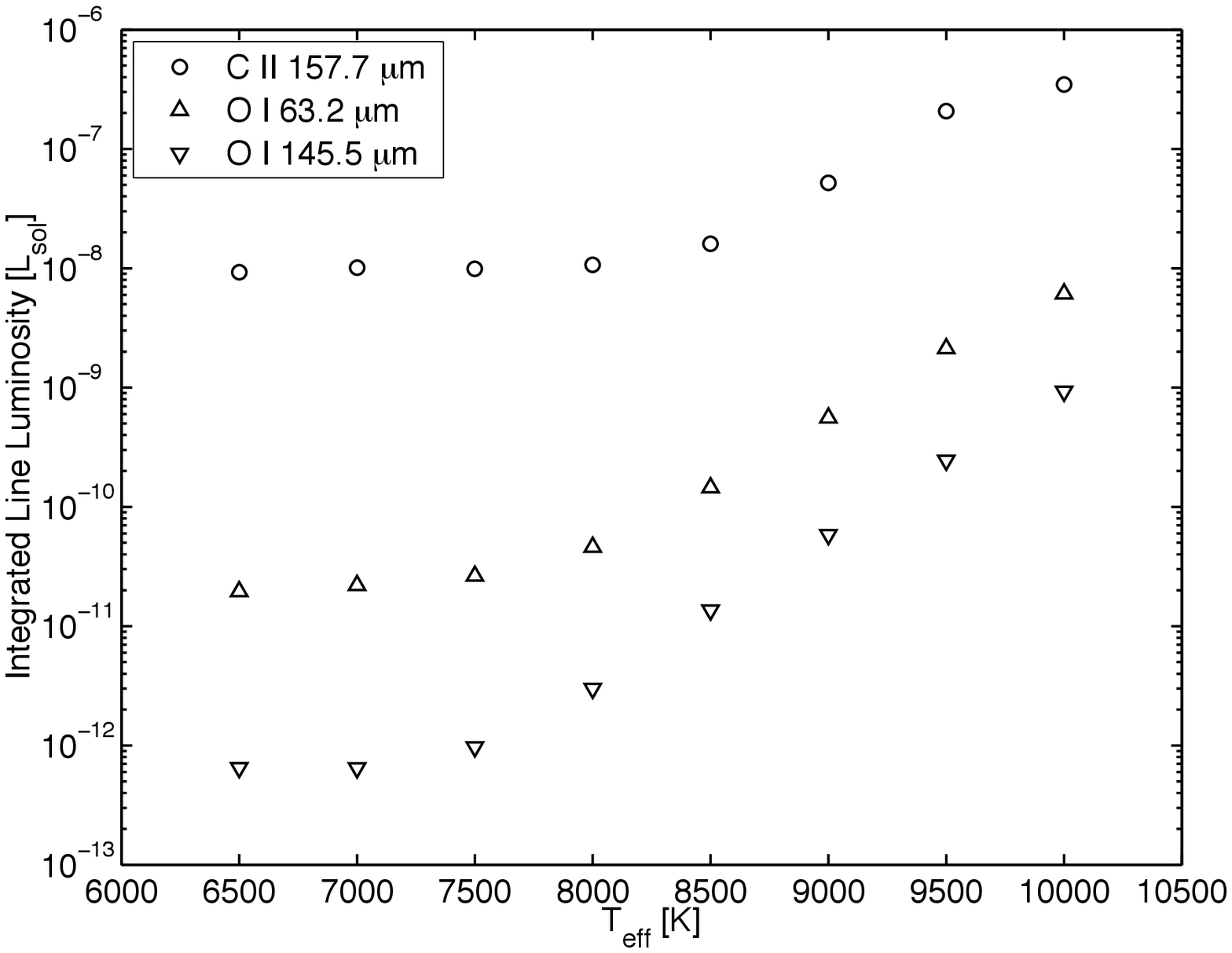} \caption{\ion{C}{2} 157.7\,$\mu$m and
      \ion{O}{1} 63.2\,$\mu$m,145.5\,$\mu$m integrated line
      luminosities as function of effective temperature
      ($T_{\mathrm{eff}}$) of the central main-sequence star. Around
      stars with $T_\mathrm{eff} \gtrsim 8\,500$\,K, the level
      population of \ion{C}{2} is dominated by radiative pumping,
      while population in \ion{O}{1} is dominated by radiative 
pumping for stars with
$T_\mathrm{eff} \gtrsim
      7\,500$\,K.}
\label{fig:Cool_vs_SpT}
\end{center}
\end{figure}

In Fig.~\ref{fig:Cool_vs_SpT}, \ion{C}{2} 157.7\,$\mu$m, \ion{O}{1}
63.2\,$\mu$m and 145.5\,$\mu$m luminosities are plotted against
stellar spectral types (represented by their respective effective
temperatures). The \ion{C}{2} 157.7\,$\mu$m line remains dominant for
all spectral types. Its luminosity is constant for $T_{\mathrm{eff}}
\leq 8\,000$\,K. Around more luminous stars the excited level of
\ion{C}{2} is populated mostly by stellar radiation (through
fluorescence of UV lines), rather than by collisions with electrons.
In this regime, the \ion{C}{2} 157.7\,$\mu$m flux is greater than the
actual cooling and it increases with $T_{\mathrm{eff}}$ until $T_{\mathrm{eff}}$
reaches $\sim 10\,000$\,K.
Relatedly, level population in the excited states of the \ion{O}{1}
atoms is dominated by radiative pumping for all stars with
$T_{\mathrm{eff}}\geq 7\,500$\,K. As a result, the line flux rises
with $T_{\mathrm{eff}}$ above 7\,500\,K.

\section{Conclusions}
\label{sec:Conclusion}

We have produced a thermal and statistical equilibrium code to
specifically address gas emission from dusty gas-poor disks around
A--F stars, motivated by the recent launch of the far-infrared
observatory
\textit{Herschel}, as well as the discovery of gas in a number of
debris disk systems.  We study the effects of, among other things,
photoelectric heating from dust grains, photoionization heating of the
gas, thermal equilibrium reached in the disk, statistical equilibrium
of atoms in the disk, and cooling by infrared atomic lines.  Details
about the code (\textsc{ontario}) are presented in the appendix.

Using this code and disk parameters that resemble the observed $\beta$\,Pic
gas disk, we have computed a range of observables, including the line
luminosities from infrared transitions, the emission and absorption of
optical and UV lines, and the column densities of metals in the disk.

As a guidance for observations to be carried out on Herschel and other
telescopes, we explore the dependence of the infrared line
luminosities on disk parameters. Our findings are summarized below:

\begin{itemize}

\item for most of the configurations surveyed, we find that 
the \ion{C}{2} 157.7\,$\mu$m line is the most luminous. The
expected flux for the $\beta$\,Pic disk is $\sim 10^{-14}$
\,erg\,s$^{-1}$\,cm$^{-2}$, which should be easily detectable by
\textit{Herschel}.

\item the \ion{O}{1} lines, initially thought to be important cooling
  lines \citep{bes07}, have underwhelming luminosities due to NLTE
  effects. Most of the luminosities in these lines arise from
  fluorescence of stellar UV photons, as opposed to from collisional
  cooling of the gas.  The flux from the \ion{O}{1} 63.2\,$\mu$m line
  will be too faint to reasonably be detected by \textit{Herschel}.

\item over the parameter range we sample, the \ion{C}{2} line 
flux scales roughly linearly with the gas density, with the carbon
elemental abundance, and rises with the effective temperature of the
star. Line fluxes also reach maximum when the dust and gas components
in the disk align radially.

\end{itemize}

We note two major caveats in our model: 

\begin{itemize}

\item in some of our models, the gas temperature rise above $5\,000$\,K. 
This is possibly an artificial result because we have not included all
relevant gas cooling mechanisms. We have performed preliminary tests
by considering other fine transition lines listed in \citep{hol89} but
do not find them of importance.  However, if this high temperature
feature does occur in real disks, it implies loss of metallic gas by
thermal evaporation.

\item we have assumed that molecular species are unimportant for the
thermal equilibrium of the gas. For cooler stars, a significant amount
of the gas may be in molecular phases (CH, CO,...) and this may impact
our predictions of infrared line luminosities.

\end{itemize}

By comparing \textsc{ontario} predictions with observations, one may
hope to infer physical properties of the debris disks, including gas
density and elemental abundances.  With the new observing windows
being opened up by \textit{Herschel},
\textit{SOFIA}, and \textit{ALMA}, we expect many new
observational results that will require interpretation by 
models such as \textsc{ontario}.

\acknowledgments KZ was funded under NSERC USRA scholarship. AB is
funded by the \textit{Swedish National Space Board} (contract
84/08:1), and YW acknowledges the NSERC support. We thank the
  referee Inga Kamp for many detailed and insightful comments, which
  helped improve the manuscript.

\appendix

\section{The \textsc{ontario} code}
\begin{figure}
\centering
   \includegraphics*[scale=0.5]{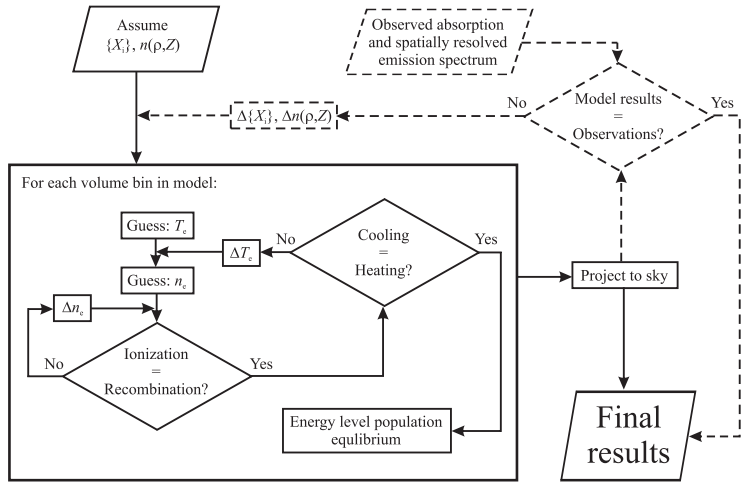}
   \caption{Model flowchart describing the \textsc{ontario} code. The dashed path
illustrates how the code is run when fitting observations.}
   \label{fig:flowchart}
\end{figure}

An overview of \textsc{ontario}'s program flow is shown in
Fig.~\ref{fig:flowchart}. \textsc{ontario} in principle 
consists of a master routine and four principal subroutines.
The four subroutines are:
\begin{enumerate}
\item \textit{Geometry solver}: defines the geometry and produces
observables from the solution. Our models are currently two
dimensional (radius and height), but could in principal be of any
geometry. The geometry solver computes the projected flux from
the defined geometry.
\item \textit{Ionization balance}: given a temperature, radiation field,
and elemental abundances, it guesses an electron density and then
iterates until the derived electron density (computed from the ionization 
of the actual elements input) agrees with the assumed electron density.
Ionization cross-sections and recombination coefficients are retrieved
from the \textsc{cloudy} code \citep{fer05}, and are restricted to
the first 30 elements (atomic numbers 1--30, hydrogen through zinc).
To estimate the radiation field, we use NextGen
stellar atmospheres \citep{hau99} with solar abundance
and main sequence surface $\log g$. No chromosphere emission is
estimated in general, but the interstellar radiation field 
\citep[from][]{wei01} and cosmic ray ionization \citep[$2 \times
10^{-17}$\,s$^{-1}$\,atom$^{-1}$, from][]{spi78} are included. 

\item \textit{Thermal balance}: given an ionization state, dust density and radiation 
field, this routine computes all relevant heating and cooling mechanisms
(as outlined in \S\,\ref{sec:thermal}) assuming an initial temperature, and then 
modifying the assumed temperature
until the heating equals the cooling. If the resulting temperature is different
from what was assumed for the ionization balance, that routine
recomputes the ionization state, which is then input to the thermal balance
routine for a new thermal balance computation.
\item \textit{Level population}: given a radiation field, an electron density and
electron temperature, the electronic energy level population is calculated using a statistical
equilibrium (SE) of all considered transitions for given elements. This routine is mostly used at 
the end, when the
physical state of the disk is already computed, as most levels and species do
not participate in the cooling of the disk. The exception is \ion{O}{1},
\ion{C}{2}, and \ion{Si}{2} where we do include a statistical equilibrium computation
already in the thermal balance computation, because of the important
cooling lines of those species. The radiative data for all atomic
species was taken from the NIST Atomic Spectra Database \citep{ral05}
and the Kurucz Atomic Spectral Line Database
\citep{kur95}. The transition data from the two sources
were combined to produce a more complete energy level set. Collisional
coefficients for the cooling \ion{C}{2} and \ion{O}{1} lines were
obtained from the Iron Project \citep{pra00} and \citet{sil01},
respectively; for \ion{C}{1}, \ion{Si}{1}, \ion{Si}{2} and \ion{S}{1} from
\citet{hol89}. Collisional data for other electron
transition lines were taken from the Iron Project
\citep{pra00} for \ion{C}{2}, \ion{Ni}{2}, \ion{Fe}{1}, \ion{Fe}{2}. At 
higher temperatures, where there were no data available, the collisional
strength was extrapolated using a first order polynomial fit in the log-log
space.  The number of levels and transitions for different elements included
in \textsc{ontario} are summarized in Table~\ref{table:bookkeeping}.
\end{enumerate}
\textsc{ontario} assumes the disk is optically thin in the continuum, 
so that simplified radiative transfer can be used for both the ionization
and thermal balance, which simplifies the computation greatly and allows
for more general geometries (compared to, e.g., \textsc{cloudy}).

Collisional strengths $\gamma$ for the electronic transition lines for
the majority of atomic species are not available from
literature. $\gamma$ is calculated by taking into account a number of
quantum effects (see, for example, \citet{zha95}) and no
simple relationship exists between the collisional strengths and the
energy associated with a transition line. Since we were only able to obtain
collisional data for the \ion{Si}{2} fine structure cooling line and not other
higher level electronic transitions, we investigated whether this
might result in incorrect determination of the gas
temperature. Results of running the code with and without \ion{C}{2}
non-cooling collisional strengths (that is, setting $q_{ij} = q_{ji} =
0$ for $i,j \neq 1,0$) for a number of bins produced no noticeable
change in the computed temperatures (within 1\,K). Therefore, all
non-cooling electron transitions can be treated as purely radiative
for gas densities representative of debris disks.

SE calculation is the most computationally expensive part of the
numerical model. An example on the extreme side is \ion{Fe}{2}, which has
790 electron energy levels and 54\,054 transitions. To solve SE for
this atomic species matrix of 54\,054 elements has to be inverted for
each bin in the grid (the actual number of levels used was reduced to 300 lower levels: the population of the higher levels is negligible, and this reduction significantly improved computation time). Furthermore, while \ion{O}{1}, \ion{C}{2} and \ion{Si}{2} have
substantially less electron energy levels (150, 191 and 143
respectively) and corresponding transitions (1048, 1347 and 927
respectively), SE for these species participating in cooling of the
gas has to be solved multiple times for each bin when recursively
computing thermal equilibrium.

To avoid a full SE computation, 2 and 3-level atom approximation for
\ion{C}{2} and \ion{O}{1} fine structure line cooling, respectively,
were previously used in similar simulations \citep{hol89,kam01}. We
used \textsc{ontario} to test the validity of this assumption for the
hotter stars sampled by our model. For \ion{C}{2} around a
$T_{\mathrm{eff}} = 10\,000$\,K star we found a factor of 4 increase
in the cooling line flux when using an ion with all the energy levels
compared to a 2-level approximation. The difference was determined to
be due to fluorescence (radiative pumping to higher levels). While the
$0\rightarrow 1$ energy transition is radiatively forbidden, ground
state electrons can get photo-excited to higher energy levels and then
spontaneously decay to the first excited level. Our results show
  that a 2-level \ion{C}{2} approximation underestimates the
  population of level 1, and consequently the $1\rightarrow 0$
  emission flux. The effect is more pronounced closer to the star
where the photo-excitation is more important. Similar results were
seen for \ion{O}{1} and \ion{Si}{2}. Furthermore, since gas densities
are substantially below $n_{\mathrm{e,crit}}$ for these species, the
effect of radiative pumping is much stronger. In fact, using a
significantly reduced number of levels for \ion{O}{1} and \ion{Si}{2}
can for some system configurations underestimate flux by more than a
factor of a 100. In addition, strong radiative pumping can
overpopulate the first excited energy state beyond its LTE level,
leading to stronger collisional de-excitation than excitation flux. In
this regime the fine-structure transition becomes a heating line,
injecting stellar energy into the gas. This effect is absent when
radiative transitions are excluded.

We further tested the effect of reducing the number of energy levels
for the cooling atoms and determined that a limited reduction did not
significantly effect the computed temperature and line fluxes. Using
10 level approximations for \ion{C}{2}, \ion{Si}{2} and \ion{O}{1} species
kept temperature close to the full-level
species results and produced similar line fluxes. However, the
decrease in computation time was significant. For example, reducing \ion{C}{2} from 191
to 10 energy levels decreases execution time by a factor of $\sim
365$ (the square of the level reduction). Therefore, we used these limited reduction
cooling species approximations in all further numerical runs.

\begin{table}[ht]
\caption{Number of energy levels and transitions included in \textsc{ontario}}
\centering
\begin{tabular}{l c c l}
\hline
Species& $\#$ of Energy Levels& $\#$ of Transitions & Refs for collisional rates.\\
\hline
\ion{C}{1}& 611 & 7597 & \citet{hol89} \\
\ion{C}{2}& 191 & 1347 & \citet{pra00} \\
\ion{O}{1}& 150 & 1048 & \citet{sil01} \\
\ion{Na}{1}& 58 & 452 & \nodata \\
\ion{Mg}{2}& 57 & 540 & \nodata  \\
\ion{Al}{1}& 213 & 798 & \nodata \\
\ion{Al}{2}& 153 & 2167 & \nodata  \\
\ion{Si}{1}& 598 & 6557 & \citet{hol89} \\
\ion{Si}{2}& 143 & 927 &  \citet{hol89} \\
\ion{S}{1}& 114 & 720 & \citet{hol89} \\
\ion{Ca}{2}& 67 & 625 & \nodata \\
\ion{Ti}{1}& 394 & 12709 & \nodata \\
\ion{Ti}{2}& 213 & 4571 & \nodata \\
\ion{Cr}{2}& 725 & 36350 & \nodata \\
\ion{Mn}{1}& 431 & 9451 & \nodata \\
\ion{Mn}{2}& 503 & 18950 & \nodata \\
\ion{Fe}{1}& 497 & 18349 & \citet{pra00} \\
\ion{Fe}{2}& 300 & 13992 & \citet{pra00} \\
\ion{Ni}{2}& 682 & 30767 & \citet{pra00} \\
\ion{Zn}{1}& 35 & 1135 & \nodata \\
\ion{Zn}{2}& 6 & 6 & \nodata \\\hline
\end{tabular}
\tablecomments{
The radiative data for all atomic
species was taken from the NIST Atomic Spectra Database \citep{ral05}
and the Kurucz Atomic Spectral Line Database
\citep{kur95}. No collisional data were available for \ion{Na}{1}, \ion{Mg}{2}, \ion{Al}{1}, \ion{Al}{2}, \ion{Ca}{2}, \ion{Ti}{1}, \ion{Ti}{2}, \ion{Cr}{2}, \ion{Mn}{1}, \ion{Mn}{2}, \ion{Zn}{1} and \ion{Zn}{2}. 
}
\tablecomments{
\edII{When computing cooling rates, only the 10 lowest energy levels were
considered for the \ion{O}{1}, \ion{C}{2} and \ion{Si}{2} species. This reduction
did not significantly affect the thermal equilibrium, but greatly
reduced computational time (see Appendix text). The full set of transition data used by
\textsc{ontario} can be retrieved from the authors upon request.}
}
\label{table:bookkeeping}
\end{table}

\end{document}